\documentclass[twocolumn,tighten,times]{aastex61}
\usepackage{natbib}
\usepackage{times}
\usepackage{color}
\usepackage{gensymb}
\usepackage{amsmath}
\usepackage{comment}

\newcommand{\msun}{\ensuremath{M_\sun}}
\newcommand{\mev}{\mbox{MeV}}

\newcommand{\kboltz}{\ensuremath{\mathrm{k_B}}}

\newcommand{\kmps}{\ensuremath{\mbox{km~s}^{-1}}}

\newcommand{\gcc}{\ensuremath{{\mbox{g~cm}}^{-3}}}
\newcommand{\mcut}{\ensuremath{M_\mathrm{cut}}}
\newcommand{\mcutX}[1]{\ensuremath{\mcut^{\mathrm{#1}}}}

\newcommand{\Abar}{\ensuremath{\overline{\mathrm{A}}}}
\newcommand{\Ye}{\ensuremath{Y_{\mathrm{e}}}}

\newcommand{\vrad}{\ensuremath{v_{\mathrm{r}}}}

\newcommand{\Ntp}{\ensuremath{N_{\mathrm{tp}}}}

\newcommand{\alp}{\ensuremath{\alpha}}
\newcommand{\eps}{\ensuremath{\epsilon}}
\newcommand{\epsi}{\ensuremath{\eps^i}}
\newcommand{\epshat}{\ensuremath{\lbrack{\eps}\rbrack}}
\newcommand{\epsihat}{\ensuremath{\lbrack{\epsi}\rbrack}}
\newcommand{\nue}{\ensuremath{\nu_{e}}}
\newcommand{\nuebar}{\ensuremath{\bar \nu_e}}
\newcommand{\numt}{\ensuremath{\nu_{\mu\tau}}}
\newcommand{\numtbar}{\ensuremath{\bar \nu_{\mu\tau}}}
\newcommand{\numu}{\ensuremath{\nu_{\mu}}}
\newcommand{\nutau}{\ensuremath{\nu_{\tau}}}
\newcommand{\numubar}{\ensuremath{\bar \nu_{\mu}}}
\newcommand{\nutaubar}{\ensuremath{\bar \nu_{\tau}}}
\newcommand{\etau}{\ensuremath{\tau^{*}}}
\newcommand{\etaumax}{\ensuremath{\etau_{\max}}}
\newcommand{\etaumin}{\ensuremath{\etau_{\min}}}

\newcommand{\dt}{\ensuremath{\delta_{t}}}
\newcommand{\dti}{\ensuremath{\dt^i}}
\newcommand{\detau}{\ensuremath{\delta_{\etau}}}
\newcommand{\detaui}{\ensuremath{\detau^i}}

\newcommand{\dNSE}{\ensuremath{\delta_{\mathrm{NSE}}}}
\newcommand{\dNSEi}{\ensuremath{\dNSE^i}}

\newcommand{\dDelr}{\ensuremath{\delta_{\Delta\mathbf{r}}}}
\newcommand{\dDelri}{\ensuremath{\dDelr^i}}

\newcommand{\nuproc}{\ensuremath{{\nu}p}-process}

\newcommand{\rproc}{\ensuremath{r}-process}

\newcommand{\phinu}{\ensuremath{\phi_\mathrm{\nu}}}
\newcommand{\Tnu}{\ensuremath{T_\mathrm{\nu}}}

\newcommand{\rfrac}[2]{{}^{#1}\!/_{#2}}
\newcommand{\e}[1]{\ensuremath{\times 10^{#1}}}

\newcommand{\isotope}[2]{\ensuremath{\mathrm{^{#2}#1}}}

\newcommand{\tpb}{\ensuremath{t_{\mathrm{pb}}}}
\newcommand{\tX}[1]{\ensuremath{t_{\mathrm{#1}}}}
\newcommand{\tfinal}{\ensuremath{t_{\mathrm{f}}}}
\newcommand{\texpl}{\ensuremath{t_{\mathrm{expl}}}}
\newcommand{\tfinalX}[1]{\ensuremath{t_{\mathrm{f-#1}}}}
\newcommand{\tNSE}{\ensuremath{t_{\mathrm{NSE}}}}
\newcommand{\tpeak}{\ensuremath{t_{\mathrm{peak}}}}
\newcommand{\tfreezeout}{\ensuremath{t_{\mathrm{fo}}}}

\newcommand{\Tfinal}{\ensuremath{T_{\mathrm{f}}}}
\newcommand{\TNSE}{\ensuremath{T_{\mathrm{NSE}}}}
\newcommand{\Tpeak}{\ensuremath{T_{\mathrm{peak}}}}

\newcommand{\rhofinal}{\ensuremath{\rho_{\mathrm{f}}}}
\newcommand{\rhopeak}{\ensuremath{\rho_{\mathrm{peak}}}}

\newcommand{\Pset}{\ensuremath{\mathbb{P}}}
\newcommand{\Pbound}{\ensuremath{\Pset_{\mathrm{bound}}}}
\newcommand{\PPNS}{\ensuremath{\Pset_{\mathrm{PNS}}}}
\newcommand{\Punb}{\ensuremath{\Pset_{\mathrm{unb}}}}
\newcommand{\Punbpos}{\ensuremath{\Pset_{+}}}
\newcommand{\Punbneg}{\ensuremath{\Pset_{-}}}
\newcommand{\Punbneghat}{\ensuremath{\lbrack{\Punbneg}\rbrack}}

\newcommand{\Mej}{\ensuremath{M_{\mathrm{ej}}}}
\newcommand{\Mbound}{\ensuremath{M_{\mathrm{bound}}}}
\newcommand{\Munb}{\ensuremath{M_{\mathrm{unb}}}}
\newcommand{\Munbpos}{\ensuremath{M_{+}}}
\newcommand{\Munbneg}{\ensuremath{M_{-}}}
\newcommand{\Munbneghat}{\ensuremath{\lbrack{\Munbneg}\rbrack}}

\newcommand{\Mtp}{\ensuremath{\Delta m}}

\newcommand{\Munbi}{\ensuremath{\Munb^{i}}}

\newcommand{\Mposi}{\ensuremath{\Munbpos^{i}}}
\newcommand{\MposX}[1]{\ensuremath{\Munbpos^{\mathrm{#1}}}}

\newcommand{\MPP}{\ensuremath{M_{\mathrm{PP}}}}
\newcommand{\MPPi}{\ensuremath{\MPP^{i}}}
\newcommand{\MPPX}[1]{\ensuremath{\MPP^{\mathrm{#1}}}}
\newcommand{\Mchim}{\ensuremath{M_{\mathrm{C}}}}
\newcommand{\Mchimi}{\ensuremath{\Mchim^{i}}}
\newcommand{\MchimX}[1]{\ensuremath{\Mchim^{\mathrm{#1}}}}

\newcommand{\eden}[1]{\ensuremath{e_{\mathrm{#1}}}}

\newcommand{\edth}{\ensuremath{\eden{th}}}
\newcommand{\edkin}{\ensuremath{\eden{kin}}}
\newcommand{\edgrav}{\ensuremath{\eden{grav}}}

\newcommand{\edtot}{\ensuremath{\eden{tot}}}

\newcommand{\PNSfull}{proto-neutron star}
\newcommand{\PNS}{proto-NS}

\newcommand{\chimera}{\textsc{Chimera}}
\newcommand{\xnet}{\textsc{XNet}}

\newcommand{\prometheusvertex}{{\sc Prometheus-Vertex}}
\newcommand{\coconutvertex}{{\sc CoCoNuT-Vertex}}
\newcommand{\castro}{{\sc Castro}}
\newcommand{\zidsa}{{Zeus+IDSA}}
\newcommand{\zelmani}{{\sc {Zelmani}}}

\newcommand{\snnet}{\texttt{sn150}}

\newcommand{\SeriesB}{B-Series}

\newcommand{\LBNL}{Nuclear Science Division, Lawrence Berkeley National Laboratory, Berkeley, CA 94720, USA}
\newcommand{\UTphys}{Department of Physics and Astronomy, University of Tennessee, Knoxville, TN 37996-1200, USA}
\newcommand{\ORNLphys}{Physics Division, Oak Ridge National Laboratory, P.O. Box 2008, Oak Ridge, TN 37831-6354, USA}
\newcommand{\JICS}{Joint Institute for Computational Sciences, Oak Ridge National Laboratory, P.O. Box 2008, Oak Ridge, TN 37831-6173, USA}
\newcommand{\NCCS}{National Center for Computational Sciences, Oak Ridge National Laboratory, P.O. Box 2008, Oak Ridge, TN 37831-6164, USA}

\submitjournal{ApJ}

\shorttitle{CCSN Nucleosynthesis using Tracer Particles}
\shortauthors{Harris et al.}

\begin{document}

\title{Implications for Post-Processing Nucleosynthesis of Core-Collapse Supernova Models with Lagrangian Particles\footnote{This manuscript has been authored by UT-Battelle, LLC under Contract No. DE-AC05-00OR22725 with the U.S. Department of Energy. The United States Government retains and the publisher, by accepting the article for publication, acknowledges that the United States Government retains a non-exclusive, paid-up, irrevocable, world-wide license to publish or reproduce the published form of this manuscript, or allow others to do so, for United States Government purposes. The Department of Energy will provide public access to these results of federally sponsored research in accordance with the DOE Public Access Plan (http://energy.gov/downloads/doe-public-access-plan).}}

\author[0000-0003-3023-7140]{J.~Austin Harris}
\affiliation{\LBNL}
\affiliation{\NCCS}

\author[0000-0002-9481-9126]{W.~Raphael Hix}
\affiliation{\ORNLphys}
\affiliation{\UTphys}

\author{Merek A.~Chertkow}
\affiliation{\UTphys}

\author{C.~T.~Lee}
\affiliation{\UTphys}

\author[0000-0002-5231-0532]{Eric J.~Lentz}
\affiliation{\UTphys}
\affiliation{\ORNLphys}
\affiliation{\JICS}

\author[0000-0002-5358-5415]{O.~E.~Bronson Messer}
\affiliation{\NCCS}
\affiliation{\ORNLphys}
\affiliation{\UTphys}

\correspondingauthor{J.~Austin Harris}
\email{harrisja@ornl.gov}

\begin{abstract}
We investigate core-collapse supernova (CCSN) nucleosynthesis with self-consistent, axisymmetric (2D) simulations performed using the neutrino hydrodynamics code \textsc{Chimera}.
Computational costs have traditionally constrained the evolution of the nuclear composition within multidimensional CCSN models to, at best, a 14-species $\alpha$-network capable of tracking only $(\alpha,\gamma)$ reactions from $^{4}$He to $^{60}$Zn.
Such a simplified network limits the ability to accurately evolve detailed composition and neutronization or calculate the nuclear energy generation rate.
Lagrangian tracer particles are commonly used to extend the nuclear network evolution by incorporating more realistic networks in post-processing nucleosynthesis calculations.
However, limitations such as poor spatial resolution of the tracer particles, inconsistent thermodynamic evolution, including misestimation of expansion timescales, and uncertain determination of the multidimensional mass-cut at the end of the simulation impose uncertainties inherent to this approach.
We present a detailed analysis of the impact of such uncertainties for four self-consistent axisymmetric CCSN models initiated from stellar metallicity, non-rotating progenitors of 12~$M_\odot$, 15~$M_\odot$, 20~$M_\odot$, and 25~$M_\odot$ and evolved with the smaller $\alpha$-network to more than 1~s after the launch of an explosion.
\end{abstract}


\section{Introduction}
\label{sec:intro}

The deaths of massive stars ($M > 8\textrm{--}10$~\msun) as core-collapse supernovae (CCSNe) are an important link in our chain of origins from the Big Bang to the present.
They are the dominant source of elements in the periodic table between oxygen and iron \citep{WoWe95,ThNoHa96}, and there is strong evidence that they are correlated with the production of half the elements heavier than iron \citep{ArSaTh04}.
Core-collapse supernovae serve both to disperse elements synthesized in massive stars during their lifetimes and to synthesize and disperse new elements themselves.

Modern multidimensional simulations by several groups, utilizing spectral neutrino transport, have successfully produced explosions for a variety of progenitors in axisymmetry (2D) though the explosion is delayed by hundreds of milliseconds compared to their non-spectral counterparts \citep{BuRaJa03,BuRaJa06,BuJaRa06,BuLiDe06,BuLiDe07,BrDiMe06,BrMeHi09b,MaJa09,SuKoTa10,TaKoSu12,MuJaMa12,MuJaHe12,BrMeHi13,MuJa14,BrLeHi16}.
Successful, fully self-consistent, spectral models have also achieved neutrino-driven explosions in three spatial dimensions \citep[3D;][]{TaKoSu12,TaKoSu14,MeJaBo15,MeJaMa15,LeBrHi15}.
Qualitatively, the 2D and 3D simulations exhibit similar explosions, dominated by a small number of rising plumes that push the stalled shock outward.
In both cases, explosions are inextricably linked to fluid instabilities that can only be properly accounted for in multidimensional simulations.
Explosions can be aided by enhanced neutrino luminosities due to fluid instabilities in the \PNSfull\ (\PNS) \citep{SmWiBa81,WiMa93}, improved neutrino-heating efficiency behind the shock as a result of fluid motions induced by convective instabilities, and the standing accretion shock instability \citep[SASI;][]{BlMeDe03,BlMe06,FoGaSc07}.
However, 3D models exhibit a tendency to be delayed when compared to their 2D counterparts \citep{HaMuWo13,MeJaMa15,LeBrHi15,JaMeSu16}.
The exception is \citet{MeJaMa15}, wherein the authors evolve a qualitatively different progenitor (zero metallicity, 9.6~\msun) than the models discussed here and the works cited above.

Asymmetries are in fact present even in the pre-supernova core, with simulations going back two decades revealing complex, potentially interacting, convective carbon, neon, oxygen and silicon burning shells above the iron core \citep{BaAr98,MeAr07b,ArMe11c}.
Until recently, the effect of such asymmetries during core collapse, bounce and into the supernova explosion have been largely ignored.
\citet{CoOt13} and \citet{MuJa15} demonstrated that artificially imposed but physically reasonable asphericities in the pre-collapse progenitor can qualitatively alter the evolution of the supernova after bounce.
To build more realistic asphericities, \citet{CoChAr15} subsequently completed the last few minutes of silicon shell burning in 3D before modeling the collapse and explosion of the star, also in 3D, with approximate neutrino transport.
Further efforts in this direction seem warranted.

Despite the fundamentally multidimensional nature of a supernova explosion from its earliest moments, relatively limited work has addressed the impact of multidimensional behavior on the nucleosynthesis.
The complexity and cost of self-consistent CCSN models (abetted by their past, frequent failures to produce explosions) has led the community to largely continue to rely on nucleosynthesis predictions for CCSNe from models using a parameterized kinetic energy piston \citep{RaHeHo02,WoHe07,MaTiHu10} or thermal energy bomb \citep{NaShSa98,MaNaNo02,UmNo08}.
We have, however, learned that multidimensional effects produce significant differences in the fraction of ejecta which experiences \alp-rich freeze-out \citep{NaShSa98,MaNaNo02} and larger ejecta velocities, characterized by metal-rich clumps \citep{KiJaHi06,HaJaMu10,ElYoFr12,WoJaMu13,WoJaMu15}.

The work of \citet{PrWoBu05}, with its combination of a neutrino-driven explosion and multidimensional fluid flow, suggested that the impact on the nucleosynthesis of departing from stratified 1D simulations was significant.
Nevertheless, very little investigation of CCSN nucleosynthesis from such models has been conducted, despite the multitude of exploding first-principles models with spectral neutrino transport in recent years.
At least in part, this curious deficit can be attributed to the prolonged times after bounce the models must be evolved in order to fully characterize the ejecta and, therefore, compute the nucleosynthesis.

Nucleosynthesis studies of electron-capture supernovae (ECSNe), which arise from the collapse of oxygen-neon cores, are more mature, as these explosions trigger and complete more rapidly than in Fe-core SNe and can be obtained even in 1D simulations.
Multi-dimensional investigations of ECSN nucleosynthesis \citep{WaJaMu11,WaJaMu13a,WaJaMu13b} using spectral neutrino transport find only modest impact from multidimensional effects, which is unsurprising given the successful 1D explosions.
Multi-dimensional Fe-core CCSN models also exhibit convective overturn near the outer \PNS\ layers, potentially with even greater affect on the nucleosynthesis and, given the necessity of multidimensionality to engender these explosions, occurs merely as a subset of other multidimensional effects.
Consequently, the lessons learned from ECSN nucleosynthesis studies, with respect to multidimensional effects, provide only modest insight into the CCSN problem.

This work aims to improve this insight by critically analyzing the sources of uncertainty in methods used to study CCSN nucleosynthesis in multidimensional models.
In a subsequent paper, we will examine the detailed differences between the nucleosynthesis in these multidimensional models and parameterized, spherically symmetric (1D) models, but first we must verify that the uncertainties achieved in these multidimensional simulations are small enough that the comparison with 1D models is meaningful.
To this end, in Section~\ref{sec:nucuncertainties}, we identify these uncertainties.
In Section~\ref{sec:methodology}, we provide a description of the relevant numerical methods and key physical approximations employed by the our radiation-hydrodynamics code (\chimera) and subsequent post-processing calculations relating to the nucleosynthesis.
In Section~\ref{sec:uncertainties}, we report our findings on the uncertainties at the end of our simulations, and in Section~\ref{sec:doneness}, we quantify how these have changed with time.
Finally, in Section~\ref{sec:summary}, we generalize our conclusions beyond the specifics of our models.

\section{Nucleosynthesis Uncertainties}
\label{sec:nucuncertainties}

Calculating the nucleosynthesis of neutrino-driven, multidimensional CCSN models removes a range of necessary assumptions related to the various approaches used to generate explosions in parameterized investigations of this nucleosynthesis.
These assumptions, though physically motivated, engender certain implicit uncertainties; foremost among these are the effects of neutrino-matter interactions and turbulent fluid flow.
For fully self-consistent simulations, examination of the nucleosynthesis affords additional observable consequences of the explosion model that can be compared to observations---these may include, for example, detailed comparisons of the spatial distribution of radioactive nuclei \citep[see, e.g.,][]{WoJaMu16}.

However, this progress to greater physical fidelity in our investigations of CCSN nucleosynthesis is not without compromise.
Many of the bomb and piston models that have provided the bulk of our understanding of CCSN nucleosynthesis in recent years utilize realistic nuclear reaction networks and continue well after the supernova ejecta reaches the surface of the star, perhaps an hour after the formation of the proto-neutron star \cite[see, e.g.,][]{WoHe07,UmNo08,ChLi13}.
To contain computational cost, multidimensional studies generally fail to include large nuclear reaction networks within their simulations.
As a result, post-processing of Lagrangian thermodynamic histories with a realistic nuclear reaction network is required to generate abundances of all of the isotopes of interest.
For Lagrangian methods, like smoothed particle hydrodynamics (SPH), these Lagrangian thermodynamic histories are simply the trajectories of the (active) particles that represent the fluid in this approach.
For Eulerian methods, passive tracer particles must be added to the conventional hydrodynamic evolution, as these grid-based methods are otherwise unable to record the histories of fluid elements, instead evolving (and recording the histories of) fluid quantities at specified locations.
Resolution is a limiting factor in any computational simulation.
While SPH simulations, by their nature, provide Lagrangian fluid elements that fully sample the hydrodynamic evolution at the intrinsic resolution of the simulation, passive tracers within Eulerian simulations do not necessarily provide similar sampling.
 
Regardless of the formalism for the hydrodynamic evolution, the use of a realistic network only in post-processing removes the natural feedback between the evolution of the nuclear composition and the thermodynamic conditions.
Thus, the accuracy of post-processing nucleosynthesis is only as good as the physical fidelity of the thermonuclear evolution included within the simulations.
In most multidimensional CCSN studies, for matter at high temperature and density, this role is served by the nuclear equation of state under the assumption of nuclear statistical equilibrium (NSE), together with neutrino transport altering the neutronization.
At lower temperatures, NSE does not apply; thus, a nuclear reaction network of some type is needed.
The simplest approach used in such models is a \emph{flashing} scheme, wherein ``Oxygen'' is converted directly to ``Silicon'' at some chosen temperature, and ``Silicon'' is similarly converted to NSE at another, higher temperature \citep[see, e.g.,][]{MaJa09}.
Of greater fidelity is an \alp-network, composed of the 13 (or 14) isotopes with equal and even proton and neutron numbers between \isotope{He}{4} and \isotope{Ni}{56} (or \isotope{Zn}{60}) and the reactions, predominantly (\alp,$\gamma$), that link them.
While an \alp-network is nearly complete for carbon burning and a reasonable approximation to oxygen burning, it does not include many of the reaction channels important for the production of iron and nickel \citep{TiHoWo00}, either by silicon burning or by the recombination of dissociated nucleons and \alp-particles, resulting in misestimation of the rate of nuclear energy release for these processes.
A further limitation of an \alp-network is the inability to track neutronization from electron and neutrino captures, so the neutronization must be monitored separately or ignored entirely.
Even when handled separately, as it is in \chimera, it is difficult to evolve the neutronization correctly in the absence of NSE without all relevant species and reaction channels available.
 
The extreme cost of fully self-consistent simulations commonly limits our most realistic CCSN models to run only several hundred milliseconds after bounce, often aborting the simulations before the nucleosynthesis is complete.
These cost-motivated compromises introduce additional sources of uncertainty in the calculation of CCSN nucleosynthesis that must be accounted for if we are to maximize the understanding gained from such studies.
Early termination of these simulations requires extrapolation of the thermodynamic history if the thermonuclear evolution is still ongoing when the simulation stops.
While a number of past studies have relied on similar extrapolations for their entire thermodynamic evolution, it is nonetheless important to constrain this uncertainty.
Early termination also makes it more challenging to determine the mass-cut, which delineates the fates of the matter within the star, and judge the composition of the ejecta.
The mass-cut does not impact all isotopes equally, the common perception \citep[see, e.g.][]{DiTi98}, based on spherically symmetric models, being that this uncertainty will most strongly impact the products of \alp-rich freeze-out; however, this question merits re-examination in a multidimensional context.

\section{Methodology}
\label{sec:methodology}

Defined by a shared set of numerical methods and physical approximations, the \chimera\ \SeriesB\ simulations originate from solar-metallicity, non-rotating progenitors calculated by \citet{WoHe07} for 12~\msun\ (B12-WH07), 15~\msun\ (B15-WH07), 20~\msun\ (B20-WH07) and 25~\msun\ (B25-WH07) stars.
These simulations were evolved with the \chimera\footnote{\url{http://www.ChimeraSN.org}} code described briefly herein, incorporating modern neutrino-matter interactions, self-consistent luminosities and neutrino spectra, and coupled nuclear burning.
\citep[For a more thorough discussion of methods in \chimera, see][]{BrLeHi16}.
The \SeriesB\ simulations extend to 1.2--1.4~seconds after bounce and were carried out in axisymmetry from the onset of collapse, with the very small post-bounce roundoff errors supplying the perturbations that seed the growth of fluid instabilities.
The long times after bounce to which the \SeriesB\ simulations were evolved make them a unique resource for investigations of the uncertainties stemming from ongoing hydrodynamic activity that can impact post-processing nucleosynthesis results, even at these late epochs.
All times reported hereafter are relative to core-bounce, or equivalently as time post-bounce (\tpb).

\chimera\ is a multidimensional, radiation-hydrodynamics code for stellar core collapse with five principal components: hydrodynamics, neutrino transport, self-gravity, a nuclear reaction network, and a nuclear equation of state (EoS).
The details most relevant to this nucleosynthesis study are summarized below.

\subsection{Radiation hydrodynamics}
\label{sec:radhyd}

Hydrodynamics is evolved via a dimensionally split, Lagrangian-plus-remap scheme with piecewise parabolic reconstruction \citep[PPMLR;][]{CoWo84} as implemented in VH1 \citep{HaBlLi12}, modified to include the consistent multi-fluid advection of \citet{PlMu99}.
The self-gravity is computed using a multi-pole expansion \citep{MuSt95}, replacing the Newtonian monopole with a general relativistic (GR) monopole \citep[][Case~A]{MaDiJa06}.
The neutrino transport solver is an improved and updated version of the multi-group (frequency), flux-limited diffusion (MGFLD) implementation of \citet{Brue85}, which uses near-complete physics, solving for four neutrino species (\nue, \nuebar, $\numt=\{\numu,\nutau\}$, $\numtbar=\{\numubar,\nutaubar\}$) while allowing for neutrino-neutrino scattering, pair exchange and other opacities.

\subsection{Nucleosynthesis in \chimera}
\label{sec:nucleosynthesis}

\SeriesB\ simulations utilize the $K = 220$~\mev\ incompressibility version of the \citet{LaSw91} EoS for nuclear-matter densities ($\rho>10^{11}$~\gcc) and an enhanced version of the \citet{Coop85} EoS for $\rho<10^{11}$~\gcc\ where nuclear statistical equilibrium (NSE) applies.
These provide a four-species representation of the chemical composition, consisting of free neutrons, free protons, \alp-particles, and a representative heavy nucleus.
To improve the fidelity of the composition of matter that may eventually become part of the ejecta, in regions of NSE where the electron fraction, \Ye, is $ \geq \rfrac{26}{56}$ (the value of $Z/A$ for \isotope{Fe}{56}), a 17-species representation of the composition is used, including free neutrons, free protons, the 14 even-Z and even-A nuclei between \isotope{He}{4} and \isotope{Zn}{60} which constitute an \alp-network, as well as \isotope{Fe}{56} to conserve \Ye.
Both the representative nucleus composition and \chimera's 17-species NSE composition provide a fairly accurate representation of the nuclear composition, as long as NSE is maintained.

For regions not in NSE, the nucleosynthesis is computed within the constraints of an \alp-network (\alp, \isotope{C}{12}--\isotope{Zn}{60}) by \xnet\footnote{\url{http://eagle.phys.utk.edu/xnet/trac/}}, a fully implicit thermonuclear reaction network code \citep[for details, see][]{HiTh99b}.
This network includes only one ``effective'' rate, triple \alp, otherwise, only rates for reactions linking explicitly defined nuclei are used.
For example, only (\alp,$\gamma$) reaction rates are included in the \alp-network, without the effective inclusion of (\alp,p) and (\alp,n) reaction rates.
In regions evolved by the reaction network from the beginning of the simulations, the initial composition is determined by mapping the abundances of free nucleons and the \alp-network species directly from the stellar progenitor to the computational grid in \chimera.
An \alp-network is of course quite limited in its ability to follow the evolution of composition where there should be a significant population of species with $N \neq Z$. 
To account for neutron-rich nuclei in the stellar progenitor, abundances of all heavy nuclei which are not part of the \alp-network are bundled into an inert, representative auxiliary nucleus to conserve \Ye.
To account for free nucleons, the abundances of free nucleons are maintained in the non-NSE region, however they do not evolve.

Naturally, zones will transition in to and out of NSE during the simulation, as matter infalls toward the core, encounters the supernova shock and is expelled. 
The decision on when to transition from the NSE composition evolution to the nuclear reaction network, or vice versa, depends on trade-offs in the the ability to accurately calculate the composition and the computational cost.
To determine whether a zone should transfer into NSE, and therefore be omitted from nuclear burning, \chimera\ applies an empirically determined linear relationship between the NSE transition temperature, \TNSE, and density:
\begin{align}
	\TNSE(\rho) = 
	\begin{cases}
		(\rho - \rho_\mathrm{L}) \left(\frac{T_\mathrm{H} - T_\mathrm{L}}{\rho_\mathrm{H} - \rho_\mathrm{L}}\right) + T_\mathrm{L} & \text{if } \rho < \rho_\mathrm{H}; \\
		T_\mathrm{H} & \text{otherwise},
	\end{cases}
\label{eq:tnse}
\end{align}
where $\rho_\mathrm{L} \equiv 5\e{7}~\gcc$, $\rho_\mathrm{H} \equiv 2\e{8}~\gcc$, $T_\mathrm{L} \equiv 5.7~\mathrm{GK}$, and $T_\mathrm{H} \equiv 6.5~\mathrm{GK}$.
This relationship is based on the need for NSE to be established, or maintained, over less than a \chimera\ timestep, otherwise, the assumption of NSE is in error.
The expression for $\TNSE(\rho)$ is based loosely on calculations of the peak temperatures required to achieve NSE in parameterized explosion simulations, starting with the one zone models of \citet{WoArCl73} and including 1D models \citep[see, e.g.,][]{ThNoYo86,ThHaNo90}.
Operationally, at the beginning of a global timestep, any non-NSE zone for which $T \geq \TNSE$ is transitioned to NSE, and the composition determined by the current temperature, density and \Ye.
Similar approaches are employed by many core collapse supernova simulations codes, though the expression for \TNSE and the means to evolve the nuclear composition before NSE is achieved differ considerably. 

The transition out of NSE is more challenging, as an initial composition must be generated for the nuclear reaction network.
In fact, we find that the details of this transition out of NSE have a significant impact on the final composition of the ejecta (see Section~\ref{sec:nsetransition}).
In \chimera, there are two cases, based on the ability to generate and evolve an accurate composition in the \alp-network.
These are formally delineated by whether the abundance of heavy nuclei would be dominated by \isotope{Ni}{56}, which is actively evolved by the network, or \isotope{Fe}{56}, which is held constant.
If the composition would include more \isotope{Ni}{56} than \isotope{Fe}{56}, effectively a $\Ye \gtrsim 0.48$, a zone which is in NSE at the beginning of a timestep will be transitioned out of NSE if $T < \TNSE - 0.2 ~\mathrm{GK}$.
In this case, the initial composition is generated by \chimera 's 17 species NSE calculation, fully populating the composition evolved by the \alp-network, plus inert abundances of neutrons, protons and \isotope{Fe}{56}.
From this detailed composition, continued compositional evolution, for example, the recombination of \alp-particles into heavy nuclei, is possible.
This is a key way in which \chimera's treatment of the transition out of NSE is an improvement over most other core collapse supernova simulations codes, which have only the 4 species composition provided by the EoS with which to seed the network.
For more neutron-rich conditions, the transition out of NSE does not occur until $T < 4.9~\mathrm{GK}$.
While this later transition runs the risk of suppressing the \alp-rich freezeout under these conditions, the poor reproduction of neutron-rich compositions by the \alp-network results in large, frozen populations of nucleons and neutron-rich heavy elements that are also unphysical.
For this reason, and because lower \Ye\ is generally associated with higher densities and hence normal (or at least only mildly \alp-rich) freezeout, the NSE composition provided by either \chimera's 17 species NSE (for $\Ye \geq \rfrac{26}{56}$) or the nuclear EOS is generally less wrong than that afforded by \chimera's included network.
Of course, the real solution to this problem is the use of a nuclear reaction network that is significantly more complete than the \alp-network, a task we are undertaking \citep{LeHiHa17}.

\subsection{Tracer particle method}
\label{sec:tpm}

We provide a brief description of the implementation of 
The tracer particle method (TPM) in \chimera\ is implemented using passive, Lagrangian tracer particles to record the thermodynamic histories of individual mass elements that are then be used for post-processing nucleosynthesis calculations \citep[see, e.g.,][]{NaHaSa97,NiKoHa06,SeRoFi10,NiTaTh15}.
Following each hydrodynamic directional sweep, the position of a tracer particle in that direction at time $t^{(n)}$, $(r^{(n)},\theta^{(n)})$, is advanced to $t^{(n+1)}$ according the simple Euler method, assuming constant velocity $(v_{r}^{(n)},v_{\theta}^{(n)})$ through the time interval $\Delta t^{(n)} = t^{(n+1)} - t^{(n)}$:
\begin{eqnarray}
	r^{(n+1)} & = & r^{(n)} + v_{r}^{(n)} \Delta t^{(n)} \,\textrm{and} \\
	\theta^{(n+1)} & = & \theta^{(n)} + \frac{v_{\theta}^{(n)}}{r} \, \Delta t^{(n)}.
\label{eq:traceradvect}
\end{eqnarray}
Physical quantities are linearly interpolated in radius to tracer particle positions from the zone-center (cell-averaged) values of the computational grid; the lone exception being the interpolation of differential neutrino number fluxes, which are defined at radial zone edges.

The tracer particles are initially distributed into \emph{rows}, radial shells uniformly-spaced in mass, beginning 0.1~\msun\ inside the edge of the progenitor's iron core.
Thus, no tracers are placed in the inner iron core, conserving the tracers for matter more likely to be ejected.
The particles within each row are placed into \emph{columns} in latitude that represent uniform volume (and hence uniform mass): $\Delta(\cos \theta) = 2/N$, where $N = 40$ particles per row for \SeriesB\ models.
For B12-WH07, $\Ntp = 4,000$~tracer particles are used, translating to 40 columns and 100 rows.
Each tracer represents $\Mtp = 1.87\e{-4}~\msun$, initially extending from $\approx$890~km to the carbon-enriched oxygen-shell at $\approx$15,000~km.
This degree of tracer resolution is common among prior nucleosynthesis studies employing the TPM \citep{NaHaSa97,PrWoBu05,NiKoHa06} and is similar to the ``medium'' resolution case of \citet{NiTaTh15}.
For more massive progenitors, more tracers are employed.
However, the number of tracers grows more slowly than the mass in the silicon and oxygen shells, resulting in lower mass resolution in the more massive models.
A description of the initial tracer particle distribution and representative mass of each tracer particle for each of the \SeriesB\ models is given in detail in Table~\ref{tab:distributions}.

\begin{deluxetable*}{lcccc}
	\tablewidth{0pt}
	\tablecolumns{5}
	\tablecaption{Initial distributions of Lagrangian tracer particles \label{tab:distributions}}
	\tablehead{
		\colhead{} & \multicolumn{4}{c}{Models} \\
		\cline{2-5} \\
		\colhead{Initial distribution} & \colhead{B12-WH07} & \colhead{B15-WH07} & \colhead{B20-WH07} & \colhead{B25-WH07}
	}
	\startdata
	Number of particles (\Ntp) & 4000 & 5000 & 6000 & 8000 \\
	Number of mass shells (rows) & 100 & 125 & 150 & 200 \\
	Particles per mass shell (columns) & 40 & 40 & 40 & 40 \\
	Mass per shell [\msun\e{-3}] & 7.472 & 11.46 & 14.18 & 13.94 \\
	Mass per particle (\Mtp) [\msun\e{-4}] & 1.868 & 2.864 & 3.545 & 3.486 \\
	\sidehead{Inner boundary}
	Inner row radius [km] & 890.8 & 1,131 & 1,385 & 1,472 \\
	Inner edge mass [\msun] & 1.203 & 1.321 & 1.424 & 1.480 \\
	Average mean mass number (\Abar) & 37.77 & 42.42 & 43.05 & 42.17 \\
	\cutinhead{Outer boundary}
	Outer row radius [km] & 15,000 & 19,456 & 19,751 & 19,751 \\
	Outer edge mass [\msun] & 3.879 & 5.476 & 3.551 & 4.269 \\
	Average mean mass number (\Abar) & 15.16 & 15.01 & 16.99 & 17.05 \\
	\enddata
\end{deluxetable*}

\subsubsection{Parameterized thermodynamic trajectories}
\label{sec:trajectories}

Purely parameterized expansion profiles have long been used to study CCSN nucleosynthesis.
\citet{FoHo64} introduced the use of adiabatic expansion following a peak temperature, \Tpeak, and peak density, \rhopeak, mimicking the passage of the shock-wave, with a characteristic expansion timescale, \etau, equal to the free-fall timescale ($446 / \sqrt{\rhopeak}$~s if \rhopeak\ is measured in \gcc).
With this assumption, the thermodynamic trajectory for a single particle may be expressed as
\begin{align}
	T(t) &= \Tpeak \exp \left(-\frac{\Delta t}{3\etau}\right),\\
	\rho(t) &= \rhopeak \exp \left(-\frac{\Delta t}{\etau}\right),
\label{eq:adiexp}
\end{align}
where $\Delta t = t - \tpeak$ is the time since the shock passage.

A number of explorations \citep[see, e.g,][]{WoArCl73,MeKrCl98,HiTh99a} have used this simple parametrized model for their entire thermodynamic evolution.
\citet{MaTiHu10} utilized both this exponential decline as well as the power-law decline that results from the homologous expansion of a uniform sphere.
\citet{NiTaTh15} also employ a power-law description of the expanding matter for simulations of magnetically-driven, rotating CCSNe soon after the early ejecta drops below NSE conditions.
\citet{PaJa09} employed an exponential decrease in temperature and density to describe the first (adiabatic) stage of a homologously expanding neutrino-driven wind before switching to a power-law model at later times, representative of reduced wind acceleration occurring after early homologous expansion.
In this case, the density and temperature decline much less steeply during the power-law phase than during exponential expansion and can be calculated from the density ($\rho_0$) and temperature ($T_0$) at the time $t_0$ when the matter is subjected to the reduced acceleration.
For $t > t_0$,
\begin{align}
	T(t) &= T_0 \left(\frac{t}{t_0}\right)^{-2/3},\\
	\rho(t) &= \rho_0 \left(\frac{t}{t_0}\right)^{-2}.
\label{eq:powerlaw}
\end{align}
Despite their efforts to account for different phases in the expansion, these purely analytic parameterizations usually begin at peak temperatures and densities and expand smoothly, failing to directly capture thermodynamic variations seen in multidimensional models.

Parameterizations like these can also be employed to extrapolate self-consistent thermodynamic trajectories from hydrodynamic simulations to much later times than the original simulations.
In that case, \emph{peak} values of temperature and density are replaced, for example in Equation~\ref{eq:adiexp}, by \emph{final} values, \Tfinal\ and \rhofinal, and $\Delta t$ becomes $t - \tfinal$, the time beyond the end of the simulation.
With the preceding hydrodynamic evolution in hand, better estimated expansion timescales are possible.
For the particle data used herein, the effective expansion timescale, \etau, is calculated by averaging the instantaneous value, $\tau = -\rho/\dot{\rho}$, during periods of expansion near the end of the simulation.
The thermodynamic histories of particles are often quite noisy, so a 25-ms wide, six-degree Savitzky-Golay smoothing filter is applied to the density profile for computation of numerical derivatives \citep{PrTeVe07}.
For the epochs we discuss in this paper, we find that the final 150~ms of the simulation provides an adequate time window for properly sampling the particle behavior.
The time window is also narrowed such that deviations in $T^3/\rho$ from the final value do not exceed 10\%.
Furthermore, we exclude outliers in the calculation of \etau\ by only considering the interquartile range of discrete values of $\tau$ over the specified time window.

The exponential behavior in temperature and density is supplemented with a more slowly declining power-law trajectory (Equation~\ref{eq:powerlaw}) after nuclear burning has reached freeze-out conditions ($T \lesssim 0.5$~GK) at time \tfreezeout.
This prevents the small temperatures that result from the exponential expansion from producing spurious results below the REACLIB lower temperature limit.
The extrapolated thermodynamic trajectory for each tracer particle combines both of these parameterizations:
\begin{align}
	T(t) &= 
	\begin{cases}
		\Tfinal \exp \left(-\frac{t-\tfinal}{3\etau}\right) & \text{if } \tfinal < t \le \tfreezeout, \\
		\Tfinal \exp \left(-\frac{\tfreezeout-\tfinal}{3\etau}\right) \left(\frac{t}{\tfreezeout}\right)^{-2/3} & \text{if } t > \tfreezeout,
	\end{cases} \\
	\rho(t) &=
	\begin{cases}
		\rhofinal \exp \left(-\frac{t-\tfinal}{\etau}\right) & \text{if } \tfinal < t \le \tfreezeout, \\
		\rhofinal \exp \left(-\frac{\tfreezeout-\tfinal}{\etau}\right) \left(\frac{t}{\tfreezeout}\right)^{-2} & \text{if } t > \tfreezeout.
	\end{cases}
\label{eq:freezeout}
\end{align}
For neutrino-induced reactions extrapolated to $t > \tfinal$, we assume that \Tnu\ remains constant and the neutrino luminosity is constant therefore $\phinu \propto 1/r(t)^{2}$, where $r(t) = r(\tfinal) + \vrad(\tfinal) (t -\tfinal)$ assuming constant radial velocity.

\subsection{Nuclear reaction network}
\label{sec:networks}

Beyond the \alp-network described in Section~\ref{sec:nucleosynthesis}, we also utilize a 150-species network, which we will refer to as the \snnet-network or just \snnet.
Ranging in atomic number from $Z = 0$ to $Z = 30$ and including all isotopes from neutron number $N = Z$ to the most neutron-rich stable isotope for each element, \snnet\ represents a first-order improvement to the \alp-network.
While not sufficient to capture more exotic nuclear processes like the \nuproc\ and \rproc, this moderately-sized reaction network encompasses a significant fraction of elemental abundances and energy-producing reactions important to the core-collapse problem, allowing proper neutronization and, when coupled to the hydrodynamics, a more accurate nuclear energy generation rate.
Reaction rates are taken from the REACLIB\footnote{\url{https://groups.nscl.msu.edu/jina/reaclib/db/}} compilation \citep{CyAmFe10} and supplemented/supplanted with $\beta$-decay rates and electron capture rates on free nucleons and heavy nuclei \citep{FuFoNe85,OdHiMu94,LaMa00}.

Neutrino capture rates on free nucleons and nuclei are also included in post-processing calculations.
For this purpose, the integrated number flux of the neutrino distribution, \phinu, is recorded for \nue\ and \nuebar\ at the tracer particles' location.
Fluxes for \numt\ and \numtbar\ are also recorded but are not included in the post-processing nucleosynthesis.
The neutrino-induced reaction cross-sections also require the neutrino temperatures, \Tnu, which we calculate by fitting the numerical neutrino distribution from the radiation-hydrodynamics simulation, $n(\epsilon_\nu)$, to a Fermi-Dirac spectrum of arbitrary degeneracy $\eta$:
\begin{equation}
	n(\epsilon_\nu) = \frac{1}{F_{2}(\eta)\Tnu^3}\frac{{\epsilon_\nu}^{2}}{\exp((\epsilon_{\nu}/\Tnu)-\eta)+1},
\label{eq:fdspec}
\end{equation}
where $F_{2}(\eta)$, the second-order Fermi-Dirac integral, is used for normalization to unity.

Post-processing calculations for ejected tracer particles which attain NSE begin at a point in time near the transition out of NSE (\tNSE), with initial conditions determined by finding the appropriate NSE composition which solves the Saha equation for the selected network and the corresponding thermodynamic conditions: $\rho(\tNSE)$, $T(\tNSE)$, and $\Ye(\tNSE)$ \citep[see, e.g.,][and references therein, for more details on the NSE equations and resulting abundances]{HiMe06,HaWoEl85}.
For tracers that never reach NSE, the initial abundances, temperature, and density are interpolated directly from the composition provided in the stellar progenitor.
We then evolve the composition along the TPM-generated thermodynamic trajectories with a standalone implementation of \xnet\ using a fully implicit integration scheme \citep{HiTh99b}.
In this way, we are able to generate more detailed nucleosynthetic yields using more realistic nuclear reaction networks.

\section{Quantifying Uncertainties}
\label{sec:uncertainties}

We quantify contributions to uncertainty in multidimensional models of CCSN nucleosynthesis relating to the determination of the multidimensional mass-cut (Section~\ref{sec:masscut}), estimation of expansion timescales for use in thermodynamic extrapolation (Section~\ref{sec:extrap}),  and limitations in the spatial (Section~\ref{sec:dxresolution}) and temporal (Section~\ref{sec:dtresolution}) resolution of the Lagrangian tracer particles.
In Section~\ref{sec:nsetransition} we discuss the impact that choices in the transition from NSE has on the composition.
We also examine some consequences of the nuclear reaction network size.
The full impact of using an \alp-network must await similar simulations with a large \emph{in situ} nuclear reaction network.
Some additional sources of uncertainty, for example, those related to the stochasticity inherent in multidimensional models, are beyond the scope of this study.

\subsection{Determination of the multidimensional mass-cut}
\label{sec:masscut}

Observations and models agree that CCSNe are highly asymmetric events driven by complex and/or turbulent fluid flows.
The implications of this inherent multidimensionality on the nucleosynthesis are lost in 1D simulations, wherein a clear distinction, the mass-cut, is easily made between matter which is ejected to the interstellar medium and that which falls back to the \PNS\ (defined hereafter as the region where $\rho > 10^{11}$~\gcc).
While the placement of the mass-cut can vary in time in a 1D model, for example, as fallback results from matter formerly expanding being decelerated by interactions with the stellar envelope, the mass-cut is nonetheless a unique mass coordinate within the progenitor star.
Extending this distinction to 2D and 3D simulations is challenging, as there is no requirement that the ejecta form a contiguous region, an implicit feature of 1D simulations.
Further, the ability for accretion and outflow to occur simultaneously makes the determination of even the initial mass-cut a much more gradual process.

Ultimately, this requires evolving a model well beyond the initial development of an explosion until such time that the downflows, which have long been cut-off from the rest of the star at the shock, cease falling onto the \PNS\ and accretion dramatically slows.
Here, the limitation in our ability to extend the models, due to the computational cost of running simulations with spectral neutrino transport, is a great impediment compared to typical parameterized models.
In this study, we have extended four self-consistent models with spectral neutrino transport using \chimera\ much further than similar models have heretofore been run in order to examine this issue.
We find that the initial multidimensional mass-cut is not yet fully resolved, despite 1.2--1.4~s of evolution after bounce and $\approx$1~s after the initiation of an explosion.

Examining the mass-cut begins by defining the ejecta.
Once the supernova shock breaks through the surface of the star, the ejecta is self-defined as the matter which propagates into the ISM.
However, a definition is needed that can be applied at much earlier epochs.
The treatment of the explosion energy faces a similar challenge of definition at these early times; thus, we mirror here the treatment of the explosion energy discussed in \citet{BrLeHi16}, which is similar to the approach used by several authors in recent years.
With contributions from specific kinetic (\edkin), thermal (\edth), and gravitational binding energy (\edgrav), the specific total energy, 
\begin{equation}
	\edtot \equiv \edkin + \edth + \edgrav,
\label{eq:edtot}
\end{equation}
defines the unbound ejecta as particles for which $\edtot > 0$.
We label this set of unbound particles as \Punb\ and the corresponding mass as \Munb.
This should not to be confused with the total ejected mass, \Mej, which also includes contributions from as yet unshocked matter, both on and above \chimera's computational grid.
Likewise, we use \Pbound\ and \Mbound\ to represent the bound matter ($\edtot < 0$) outside of the \PNS.
Ideally, all tracer particles in \Punb\ would constitute a portion of the observed ejecta.
However, due to the work required to lift the stellar envelope out of the star's gravitational well, \Punb\ is likely an overestimate of particles that would ultimately be ejected.

The free-fall time, $t_\mathrm{ff} = \sqrt{3\pi/(32G\rho)}$~s \citep{KiWe90}, provides a rough estimate of a lower-bound for the additional time needed to establish the initial multidimensional mass-cut (i.e. $\delta t_\mathrm{masscut} \gtrsim t_\mathrm{ff}$).
In addition to the free-fall time, each parcel of matter represented by a particle is subject to additional neutrino heating as it nears the \PNS\ and may be re-ejected.
Letting $\rho$ be the average density of all marginally unbound ($|\edgrav|\le\edkin+\edth\le1.5|\edgrav|$) particles with relatively small radial velocities ($\vrad < 1\e{3}~\kmps$), we estimate lower-bounds for $\tfinal + \delta t_\mathrm{masscut}$ as 6.50~s, 4.29~s, 4.37~s, and 4.32~s for B12-WH07, B15-WH07, B20-WH07, and B25-WH07, respectively.

\begin{figure}
	\centering
	\includegraphics[width=\columnwidth,clip]{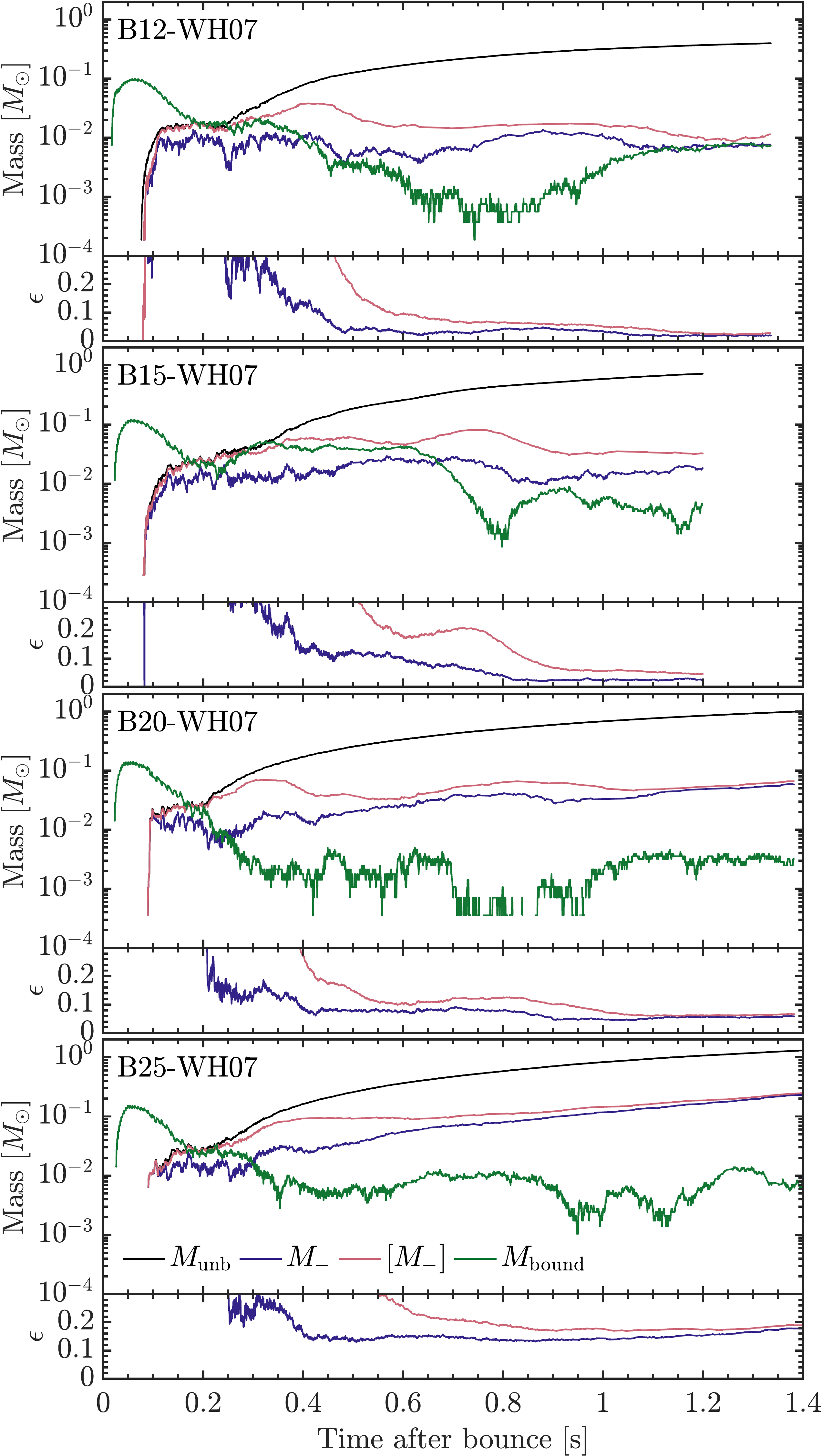}
	\caption{\label{fig:bseriesmass}
		Top panels: Total mass represented by particles in \Punb\ (black lines), \Punbneg\ (blue lines), \Punbneghat\ (red lines), and \Pbound\ (green lines) for each \SeriesB\ simulation.
		Gaps in the line for \Mbound\ in B20-WH07 indicate zero mass and are a result of limited tracer particle resolution (see Section~\ref{sec:dxresolution}).
		Bottom panels: Fraction of \Munb\ represented by particles in \Punbneg\ and \Punbneghat.
	}
\end{figure}

A further complication is that some particles in \Punb\ have negative radial velocities and, thus, increasing temperature and pressure.
Therefore, they cannot be reliably extrapolated for post-processing nucleosynthesis.
For this reason, it is helpful to categorize \Punb\ by radial velocity, which we label as \Punbpos\ and \Punbneg\ for $\vrad > 0$ and $\vrad < 0$, respectively.
We find that the ultimate fates of particles in \Punbneg\ as either ejecta or part of the \PNS\ (\PPNS) are often unknowable at the end of the simulation.
Consequently, the mass represented by \Punbneg, \Munbneg, is one indication of uncertainty in the total ejecta mass.
Therefore, trends in $\Munbneg(t)$ partially characterize the duration of time we must evolve a model in order to keep this uncertainty manageable as illustrated in Figure~\ref{fig:bseriesmass} (red lines).
Total unbound mass (\Munb, black lines) gradually increases as the shock lifts the envelope.
\Mbound\ (green lines) remains stubbornly above 0.001~\msun\ at the end of each model, reflecting the ongoing accretion evident in these models.
\Munbneg\ (blue lines) still exhibits significant growth at the end of B20-WH07 and B25-WH07, while B12-WH07 and, possibly, B15-WH07 have leveled out near 0.01~\msun.
We can attribute this behavior to an immature explosion in B25-WH07 and a weak explosion in B20-WH07.
Across all four models, the final values of \Munbneg\ exhibit a monotonic increase with progenitor mass from 0.008~\msun\ in B12-WH07 to 0.2~\msun\ in B25-WH07.

\begin{figure}
	\centering
	\includegraphics[width=\columnwidth,clip]{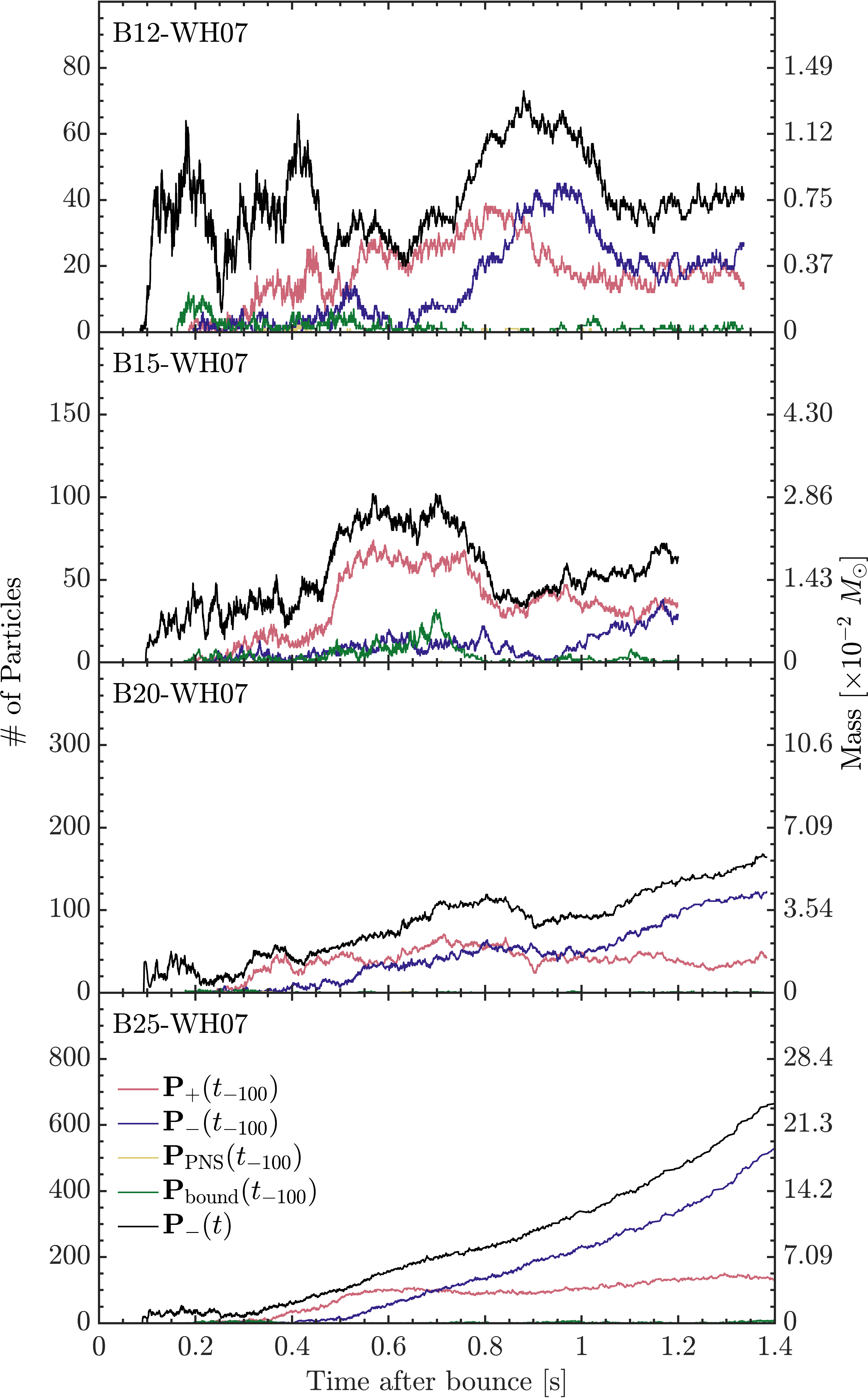}
	\caption{\label{fig:bseriesnegvr100}
		Fate classification for particles changes as the simulation progresses.
		Shown in black are the number of particles in $\Punbneg(t)$ and the corresponding mass, $\Munbneg(t)$ as a function of time $t$.
		Colored lines show the same particles' classifications at an offset time, $\tX{-100} = t - 100$~ms.
	}
\end{figure}

For B12-WH07, \Munbneg\ is relatively unchanged in the 100~ms prior to the end of the simulation, defined as \tfinalX{100}.
However, closer inspection reveals that, while the total number of particles in \Punbneg\ tracks the behavior of \Munbneg, the individual particles in \Punbneg\ are changing as convective flows and shear move particles between \Punbneg\ and \Punbpos.
This variability of \vrad\ for \Punb\ near the end of the simulation complicates the determination of a particle as being representative of the ejected matter.
In Figure~\ref{fig:bseriesnegvr100}, we illustrate this phenomenon by plotting the various fates of the particles in $\Punbneg(t)$ (black line) at $t - 100$~ms.

Ideally, $\Punbneg(t)$ and $\Punbneg(t-100~\mathrm{ms})$ would be identical, indicating a consistent determination of the particles' fates.
However, since there is a persistent fraction of particles in $\Punbneg(t)$ that are being classified as ejecta only 100~ms in the past ($\Punbpos(t-100~\mathrm{ms})$; red line), we identify all particles that were in \Punbneg\ at any time in the last 100~ms of the simulation, which we label \Punbneghat, as having indeterminate fates and contributing to a better estimated uncertainty in the ejecta mass.
We use this type of nomenclature hereafter to refer to quantities derived from \Punbneghat\ (e.g. \Munbneghat).
For B12-WH07, we see that \Munbneghat\ is $\sim$50\% larger than \Munbneg, reflecting that a significant portion of \Punbneg\ is undergoing substantial variations in radial velocity over the past 100~ms.
For B15-WH07, $\Munbneghat/\Munbneg = 1.78$ is relatively even larger, indicating more exchange of inwardly and outwardly moving particles.
Intriguingly, the much larger and still growing number of particles in \Punbneg\ in B20-WH07 and B25-WH07 are not undergoing similar variations, with \Munbneghat\ only 13\% and 7\% larger than \Munbneg, respectively.

\begin{figure}
	\centering
	\includegraphics[width=\columnwidth,clip]{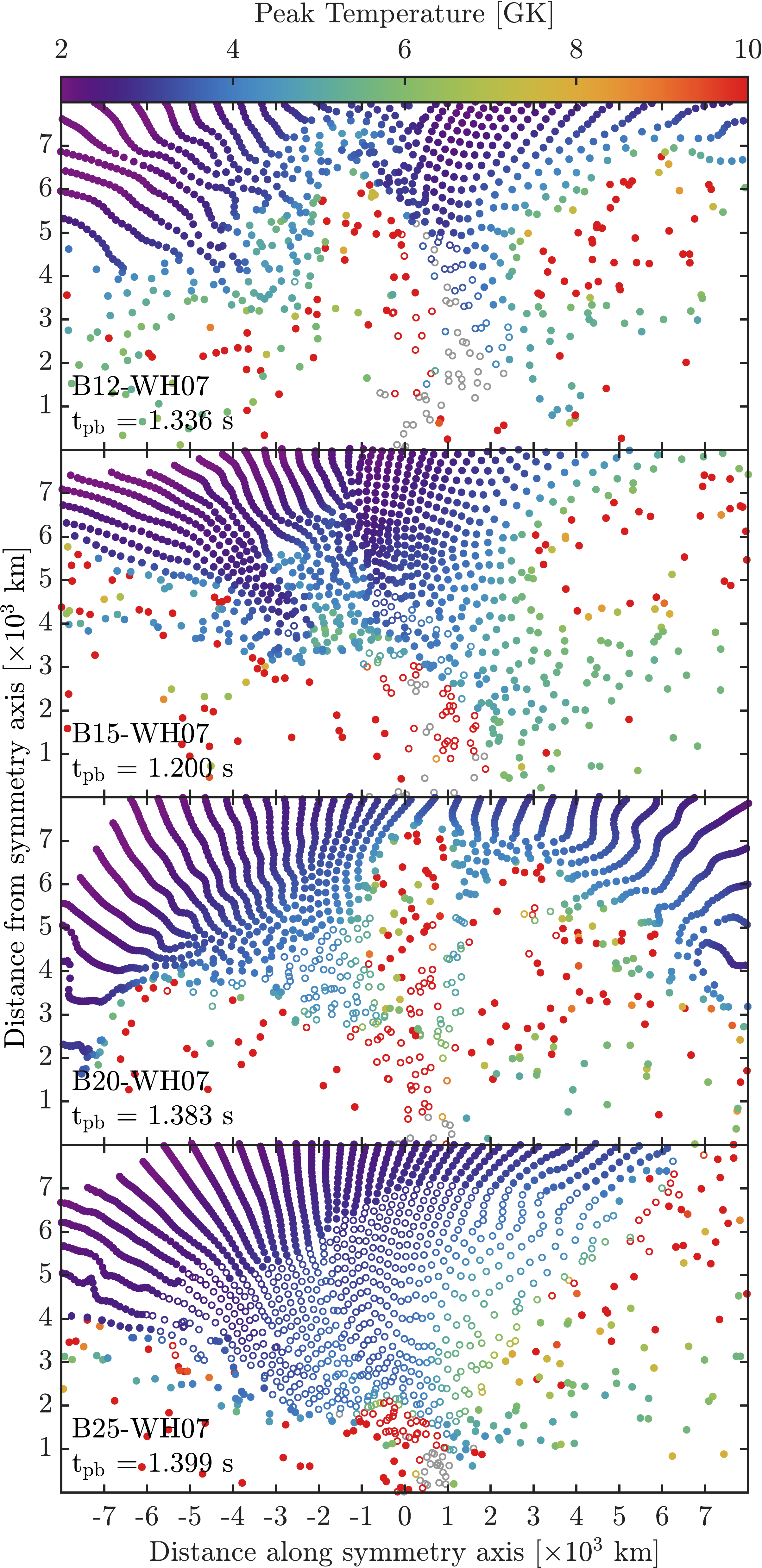}
	\caption{\label{fig:bseriespeaktempfinal}
		For each \SeriesB\ simulation, shocked tracer particles in $\Punbpos(\tfinal)$ (solid, colored circles), $\Punbneg(\tfinal)$ (open, colored circles), $\Pbound(\tfinal)$ (open, gray circles), and those located in the \PNS\ (solid, gray circles), placed at their locations at \tfinal\ and, if unbound, colored by \Tpeak.
	}
\end{figure}

The fates and peak temperatures, \Tpeak, for shocked particles in the inner 8,000~km of each of the \SeriesB\ models are shown at \tfinal\ in Figure~\ref{fig:bseriespeaktempfinal}.
The shock has generally moved beyond this grid, except in the upper left corners for B20-WH07 and B25-WH07, wherein the absence of shocked tracers is visible.
Downflows of bound matter (\Pbound; open, gray circles) are clearly visible, especially in B12-WH07 and B15-WH07, dragging unbound but now infalling particles (\Punbneg; open, colored circles).
The much larger mass represented by \Punbneg\ particles, \Munbneg, for B20-WH07 and B25-WH07, illustrated in Figure~\ref{fig:bseriesmass}, is plainly visible in the open, colored circles of Figure~\ref{fig:bseriespeaktempfinal}, especially for B25-WH07.
Clearly, in both B20-WH07 and B25-WH07, the shock in the equatorial regions has been less successful in pushing the envelope outward, suggesting significant accretion will continue for a considerable time.

\begin{figure}
	\centering
	\includegraphics[width=\columnwidth,clip]{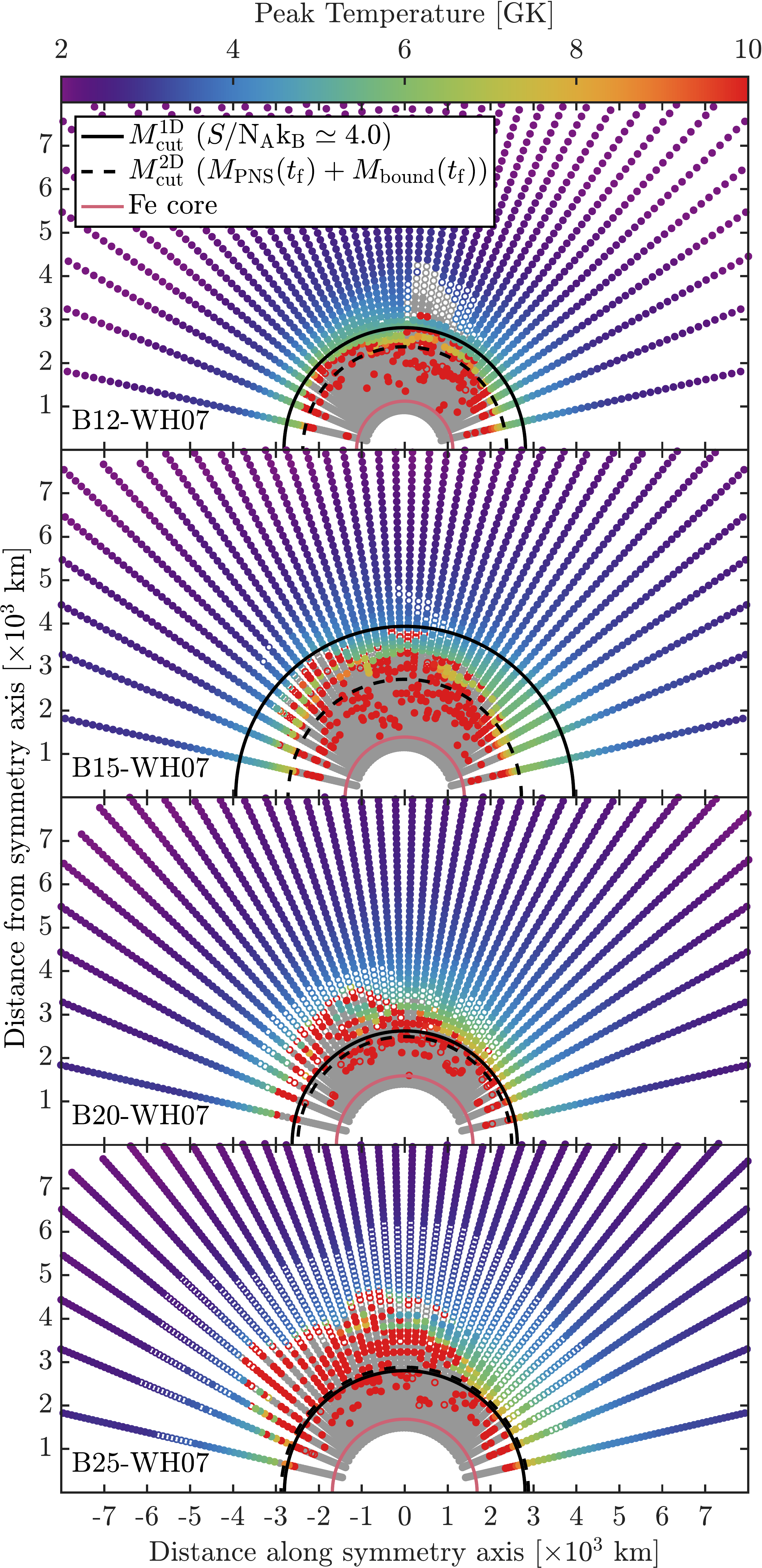}
	\caption{\label{fig:bseriespeaktempinit}
		Peak temperatures and fates for particles at their initial locations presented as described in Figure~\ref{fig:bseriespeaktempfinal} for each \SeriesB\ simulation.
		A spherical representation of the two-dimensional mass-cut (\mcutX{2D}; dashed, black line) is plotted at an enclosed mass equal to the combined \PNS\ and bound mass.
		The outer edge of the iron core (red line) and the initial location of the 1D mass-cut (\mcutX{1D}; solid, black line) are also shown.
	}
\end{figure}

Since the tracers are a Lagrangian representation of the matter, we can trace them back in time to their original positions.
Figure~\ref{fig:bseriespeaktempinit} shows the same quantities as Figure~\ref{fig:bseriespeaktempfinal}, but with the tracers positioned at their initial locations, revealing the fate of tracers as a function of their starting point in the progenitor.
These same progenitor models were exploded by \citet{WoHe07} using a parameterized piston positioned at the $S/N_{A}\kboltz = 4.0$ isoentropy contour.
This represents the mass-cut in these parameterized models, \mcutX{1D}, and generally lies at the inner boundary of the oxygen-burning shell (approximately 2,800~km, 3,900~km, 2,600~km, and 2,800~km for the 12~\msun, 15~\msun, 20~\msun, and 25~\msun\ \SeriesB\ models, respectively).
The location of \mcutX{1D} typically lies farther from the iron-core than its 2D counterpart, \mcutX{2D}, calculated as the combined mass of \PNS\ material and bound material at \tfinal.
It is important to note, however, that while a spherical mass-cut can be constructed for 2D models, it is not a good representation of the fate of individual mass elements.
While detailed analysis of the differences between spherical, parameterized models and multidimensional, neutrino-driven models will be presented in \citet{HaHiCh17}, the presence of bound particles originating above the spherical mass-cut, and unbound particles below, illustrates the fundamental difference in a multidimensional mass-cut.
Intriguingly, a significant number of ejected tracers originate in the relatively dense silicon shell, some nearly to the edge of the iron core, which has ramifications for the nucleosynthesis.
Of more immediate interest is the identification of the currently accreting matter.

In B12-WH07, B15-WH07, and B20-WH07, \Punbneg\ (open, colored circles) covers a small, isolated region of the initial particle distribution (see Figure~\ref{fig:bseriespeaktempinit}), though there is a trend to wider regions in latitude with increasing mass.
These particles represent an uncertainty in the identification of the ejecta (i.e. mass-cut) and originate 3,000--4,000~km from the center and have reached peak temperatures of 3--5~GK.
From this, we can infer that the explosive burning products of a composition originally consisting of \isotope{O}{16} and \isotope{Si}{28} are most susceptible to the mass-cut.
This is in contrast to the usual assumption that products of \alp-rich freeze-out are most susceptible to this determination.
The characteristics of \Punbneg\ in B25-WH07 are much more extreme than that of the other three models.
In the case of the 25~\msun\ model, the uncertainty in the identification of the ejecta is not confined to a small region in the initial particle distribution, but nearly extends from pole-to-pole in latitude.

\begin{figure*}
	\centering
	\includegraphics[width=\textwidth,clip]{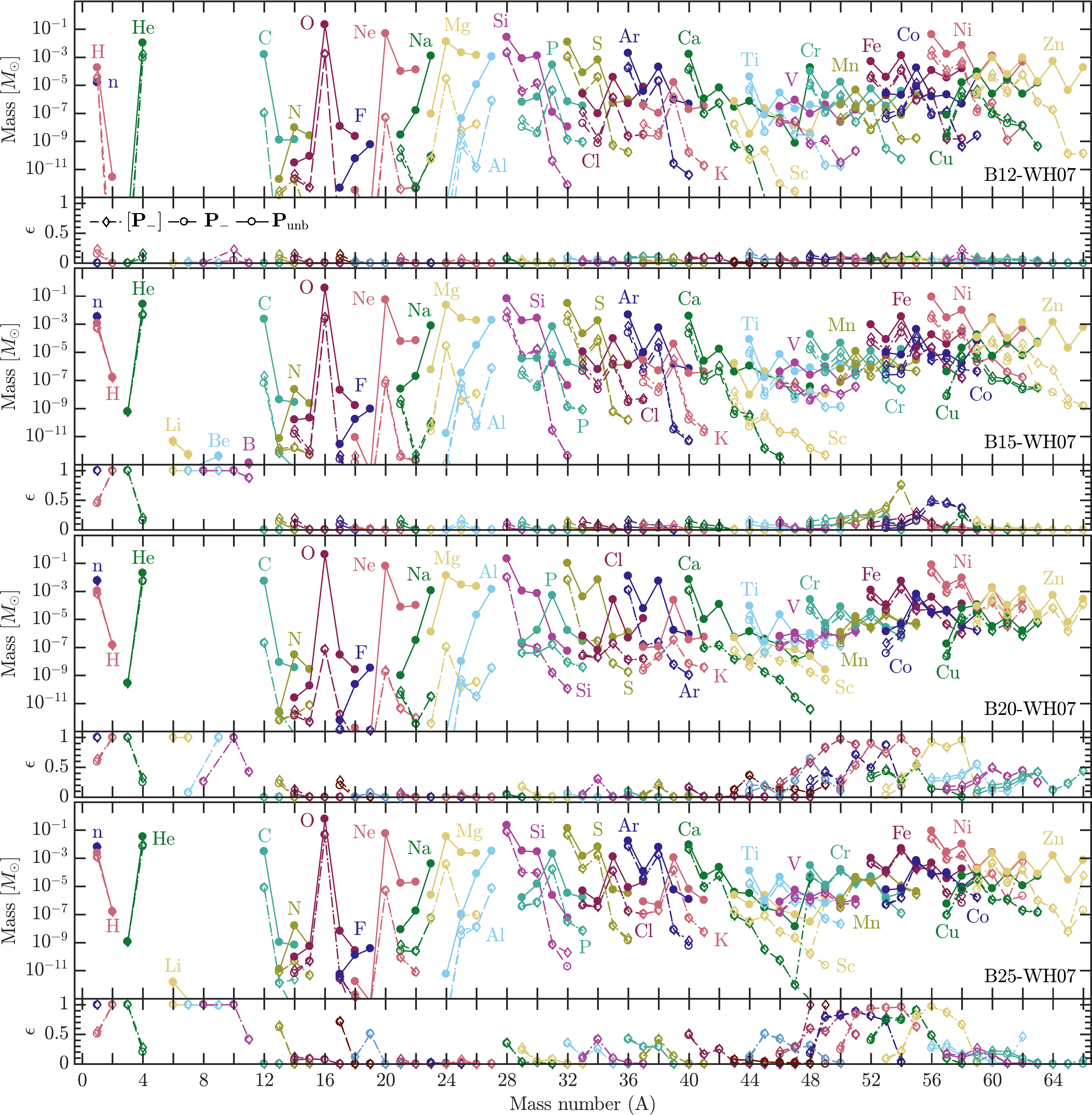}
	\caption{\label{fig:bseriesmasscut}
		Total unbound mass of individual isotopes represented by \Punb\, ($\Munbi(\tfinal)$; solid lines) and fractions thereof represented by \Punbneg\ (\epsi; dashed lines) and \Punbneghat\ (\epsihat; dashed-dotted lines) after post-processing with \snnet\ and $\TNSE = 8$~GK for each \SeriesB\ simulation.
	}
\end{figure*}

We quantify these uncertainties in all four models in Figure~\ref{fig:bseriesmasscut} by comparing the total unbound mass of each isotope after post-processing with \snnet\ to that represented by \Punbneg\ and \Punbneghat.
The values of $\eps \equiv \Munbneg / \Munb$ and $\epshat \equiv \Munbneghat / \Munb$ for each model, as well as \epsihat\ for \isotope{He}{4}, \isotope{Si}{28}, \isotope{Ti}{44}, and \isotope{Ni}{56}, are given in Table~\ref{tab:uncertainties}.
The larger disagreement of \eps\ relative to \epshat\ in B12-WH07 and B15-WH07 compared to the less evolved B20-WH07 and B25-WH07 models relates to the different behavior of \Munbneg\ and \Munbneghat\ exhibited in B20-WH07 and B25-WH07, and is a testament to the challenge of running simulations sufficiently past bounce such that the multidimensional mass-cut is truly resolved.

\begin{deluxetable*}{lccccc}
	\tabletypesize{\scriptsize}
	\tablewidth{0pt}
	\tablecolumns{6}
	\tablecaption{Nucleosynthesis uncertainties \label{tab:uncertainties}}
	\tablehead{
		\colhead{} & \colhead{} & \multicolumn{4}{c}{Models} \\
		\cline{3-6} \\
		\colhead{} & \colhead{} & \colhead{B12-WH07} & \colhead{B15-WH07} & \colhead{B20-WH07} & \colhead{B25-WH07}
	}
	\startdata
	\tpb\ at simulation end (\tfinal) [s] & & 1.336 & 1.200 & 1.383 & 1.399 \\
	\sidehead{Masses at \tfinal}
	Off-grid [\msun]
	& & 8.829 & 10.04 & 12.32 & 11.26 \\
	Unshocked (on grid) [\msun]
	& & 0.1900 & 0.3600 & 0.7852 & 1.368 \\
	\Munb\ [\msun]
	& & 0.3964 & 0.7117 & 0.9919 & 1.284 \\
	\Munbpos\ [\msun]
	& & 0.3893 & 0.6937 & 0.9334 & 1.052 \\
	\Munbneg\ [\msun\e{-2}]
	& & 0.7659 & 1.804 & 5.849 & 23.14 \\
	\Munbneghat\ [\msun\e{-2}]
	& & 1.139 & 3.208 & 6.594 & 24.71 \\
	\sidehead{Combined uncertainties at \tfinal}
	$\Munbneg / \Munb$ (Section~\ref{sec:masscut})    & & 0.0193 & 0.0254 & 0.0590 & 0.1803 \\
	$\Munbneghat / \Munb$ (Section~\ref{sec:masscut}) & & 0.0287 & 0.0451 & 0.0665 & 0.1926 \\
	\\
	$||r_{\detau}||$ (Section~\ref{sec:extrap})       & & 0.0043 & 0.0069 & 0.0101 & 0.0119 \\
	$||r_{\dDelr}||$ (Section~\ref{sec:dxresolution}) & & 0.6192 & 0.7117 & 0.7767 & 0.7086 \\
	$||r_{\dNSE}||$ (Section~\ref{sec:nsetransition}) & & 0.1083 & 0.1134 & 0.0631 & 0.0770 \\
	\sidehead{Individual isotope uncertainties at \tfinal}
	\epsihat\ (\snnet) (Section~\ref{sec:masscut})
	& \isotope{He}{4}  & 0.1607 & 0.2070 & 0.3108 & 0.2678 \\
	& \isotope{Si}{28} & 0.0863 & 0.1151 & 0.0502 & 0.3670 \\
	& \isotope{Ti}{44} & 0.1722 & 0.1274 & 0.1956 & 0.1474 \\
	& \isotope{Ni}{56} & 0.0701 & 0.0490 & 0.3239 & 0.3040 \\
	\\
	\detaui\ (\snnet) (Section~\ref{sec:extrap})
	& \isotope{Ti}{44} & 0.0272 & 0.0618 & 0.1000 & 0.0213 \\
	\\
	\dDelri\ (\alp) (Section~\ref{sec:dxresolution})
	& \isotope{He}{4}  & 0.0976 & 0.1762 & 0.1176 & 0.1600 \\
	& \isotope{Si}{28} & 0.0248 & 0.0059 & 0.0117 & 0.0083 \\
	& \isotope{Ti}{44} & 0.9383 & 0.9345 & 1.0703 & 1.0520 \\
	& \isotope{Ni}{56} & 0.0121 & 0.0130 & 0.0247 & 0.0249 \\
	\\
	\dNSEi\ (\snnet) (Section~\ref{sec:nsetransition})
	& \isotope{He}{4}  & 0.2495 & 0.2682 & 0.3108 & 0.2236\\
	& \isotope{Si}{28} & 0.0000 & 0.0013 & 0.0502 & 0.0023 \\
	& \isotope{Ti}{44} & 0.5531 & 0.5701 & 0.4101 & 0.4526 \\
	& \isotope{Ni}{56} & 0.0350 & 0.0462 & 0.0437 & 0.0711 \\
	\enddata
\end{deluxetable*}

As illustrated by Figure~\ref{fig:bseriesmasscut}, the effect of the indeterminate mass-cut in B12-WH07 is most prevalent for $A \geq 28$, and $\epsihat \lesssim 0.2$.
The impact on the production of these species can be understood by considering the region of the star where the mass-cut has not yet been determined, illustrated by the open, colored circles in Figure~\ref{fig:bseriespeaktempfinal}.
For B12-WH07, this region is confined to the inner 5,000~km of the star around a cut-off downflow rich in \isotope{Si}{28} which continues to accrete onto the \PNS\ long after the development of the explosion.
This region is similarly confined for both B15-WH07 and B20-WH07, though characterized by more matter which has had its initial composition effectively reset by attaining NSE, evidenced by greater uncertainties for iron-group isotopes.
This trend also appears in Figure~\ref{fig:bseriesmasscut}, where \epsi\ in B15-WH07, B20-WH07, and B25-WH07 is characteristically different from B12-WH07 for $A > 48$, suggesting that the isotopes of Cr, Mn, Fe, and Co with values of $Z/A$ closest to $\Ye(\tNSE)$ of the ejected material are especially susceptible to the partially-resolved, multidimensional mass-cut.
In Figure~\ref{fig:bseriespeaktempfinal}, \Punbneg\ in B25-WH07 spans a much larger region of the star, not only the result of a deeper gravitational well, but also a consequence of an explosion that has been less successful in lifting the marginally-unbound, equatorial downflow \citep[see animated Figure~2 in][]{BrLeHi16}.

\subsection{Thermodynamic extrapolation}
\label{sec:extrap}

\begin{figure}
	\centering
	\includegraphics[width=\columnwidth,clip]{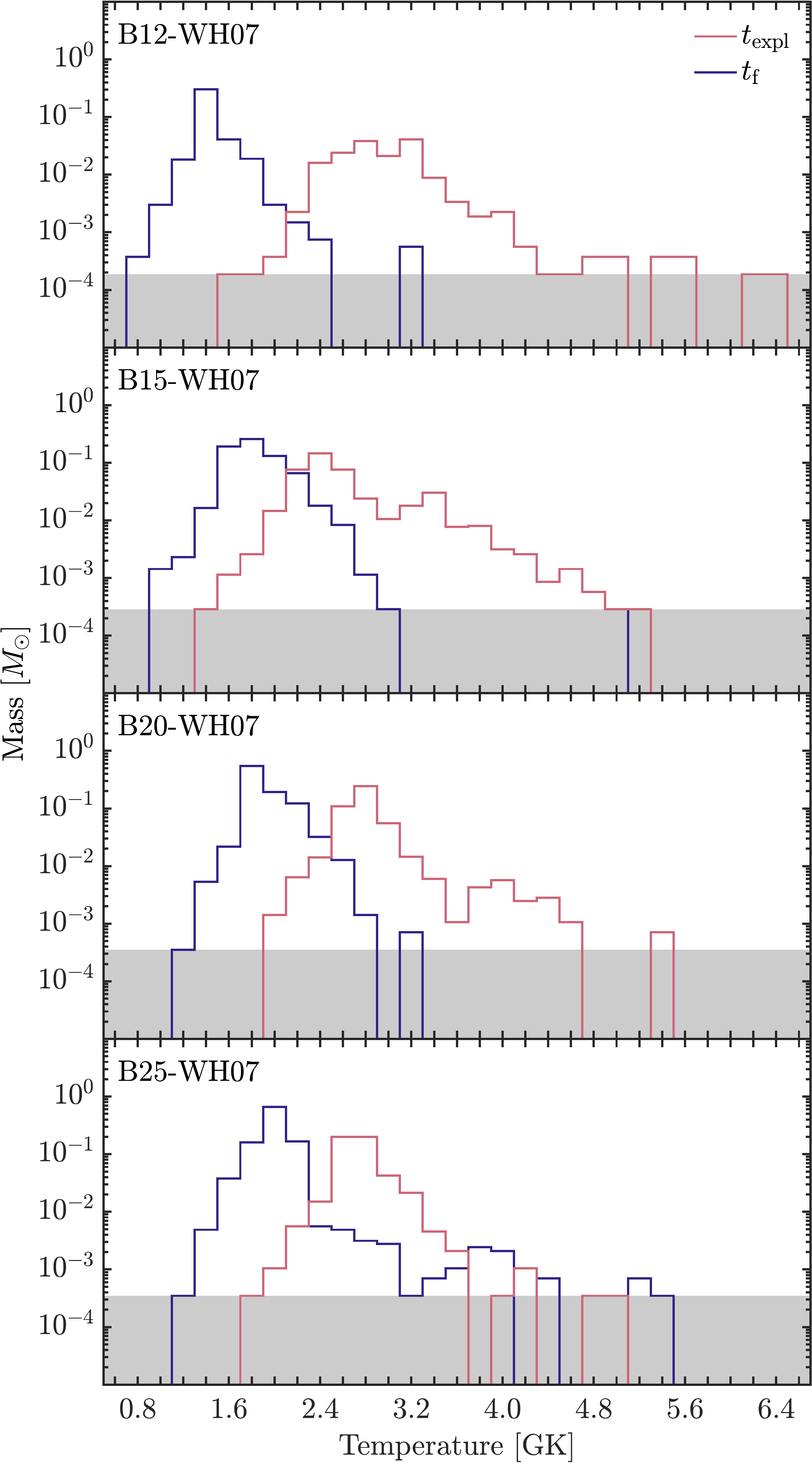}
	\caption{\label{fig:histtemp}
		\SeriesB\ models' distributions of \Munbpos\ at \tfinal\ (blue) and \texpl\ (red) in bins of size $\Delta T = 0.2$~GK.
		The mass represented by one Lagrangian tracer particle is given by the shaded region.
	}
\end{figure}

The distribution of particle temperatures at \tfinal\ in each \SeriesB\ simulation, \Tfinal, is shown in Figure~\ref{fig:histtemp} (red) for $\Punbpos(\tfinal)$.
In the case of the relatively less-evolved B25-WH07 \citep[see Figure~12 in][]{BrLeHi16}, $\approx$0.11~\msun\ of the ejecta continues to experience explosive burning ($\Tfinal \gtrsim 3$~GK), despite 1.4~s of evolution after core-bounce.
For the other three models, the nuclear reactions which account for the bulk of the nickel production in CCSN ejecta have ceased.
However, secondary nuclear burning processes will continue to alter the abundance distribution until the matter freezes out \citep{WoMaWi94}, and proton and neutron captures will continue until the temperature of the ejecta falls below $\approx$0.5~GK \citep{FrHaLi06}, particularly in proton-rich ejecta.
Short of being able to extend the simulations to this freeze-out temperature, we must extrapolate the thermodynamic conditions using the method described in Section~\ref{sec:trajectories}.

\begin{figure}
	\centering
	\includegraphics[width=\columnwidth,clip]{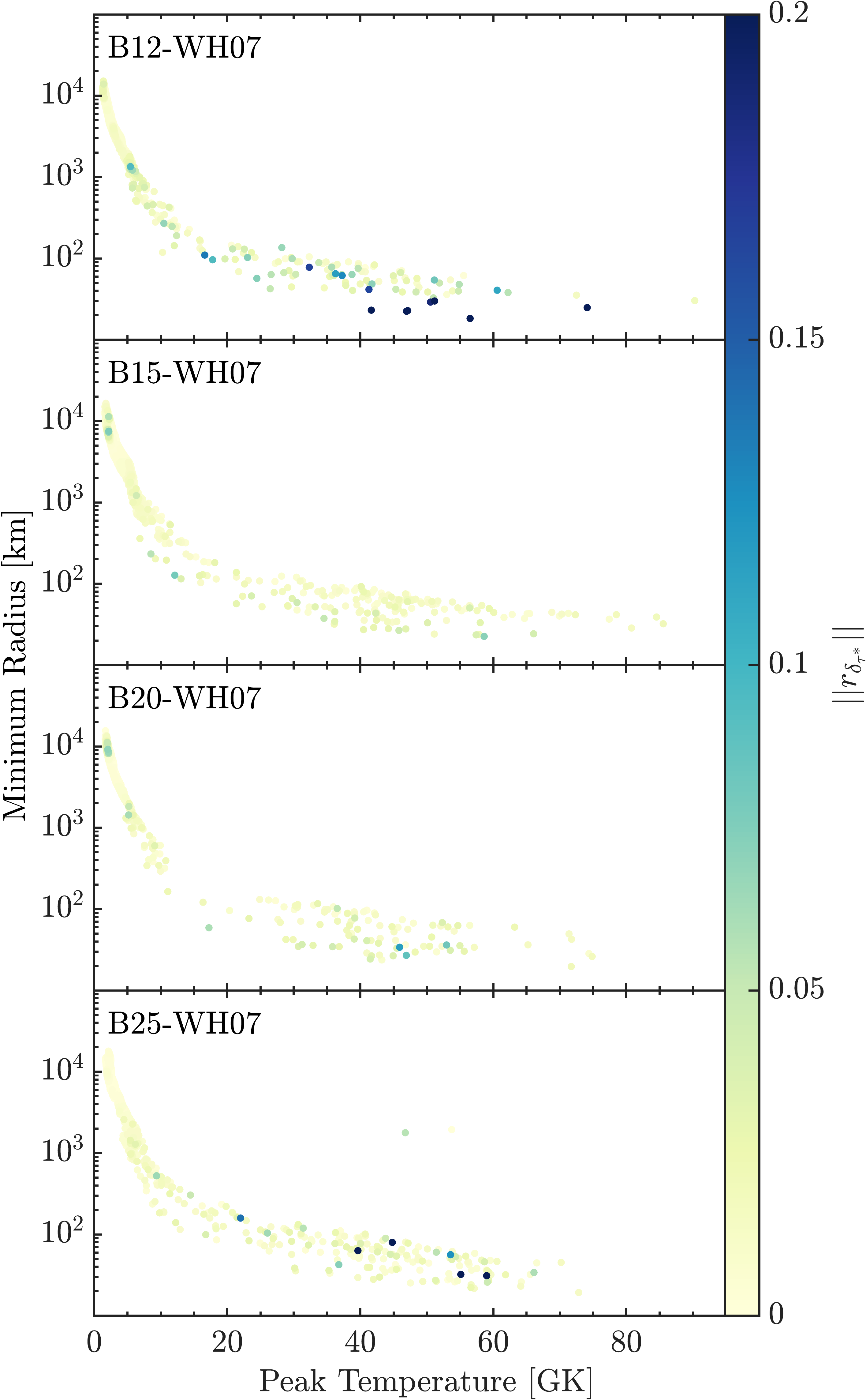}
	\caption{\label{fig:rdtau}
		Tracer particles in $\Punbpos(\tfinal)$ for each \SeriesB\ simulation and positioned according to \Tpeak\ and minimum radius and colored by the residual $||r_{\detau}||$ (Equation~\ref{eq:rdetau}).
	}
\end{figure}

\begin{figure}
	\centering
	\includegraphics[width=\columnwidth,clip]{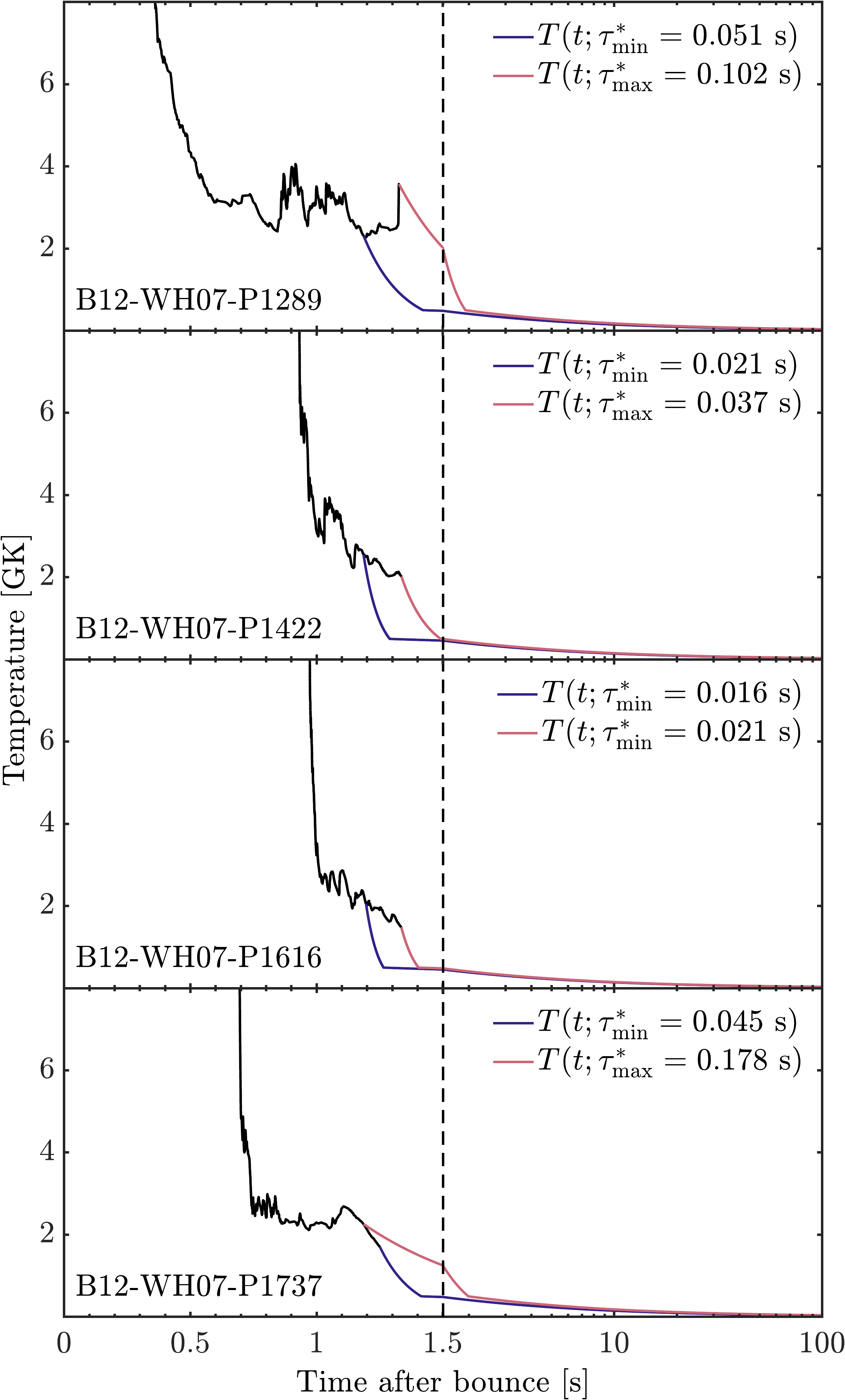}
	\caption{\label{fig:b12extrap}
		Temperature profiles and extrapolations to 100~s for B12-WH07-P1289, B12-WH07-P1422, B12-WH07-P1616, and B12-WH07-P1737, calculated using \etaumax\ (blue line) and \etaumin\ (red line).
		The dashed line indicates the switch to a logarithmic scale for the time axis.
	}
\end{figure}

\begin{figure}
	\centering
	\includegraphics[width=\columnwidth,clip]{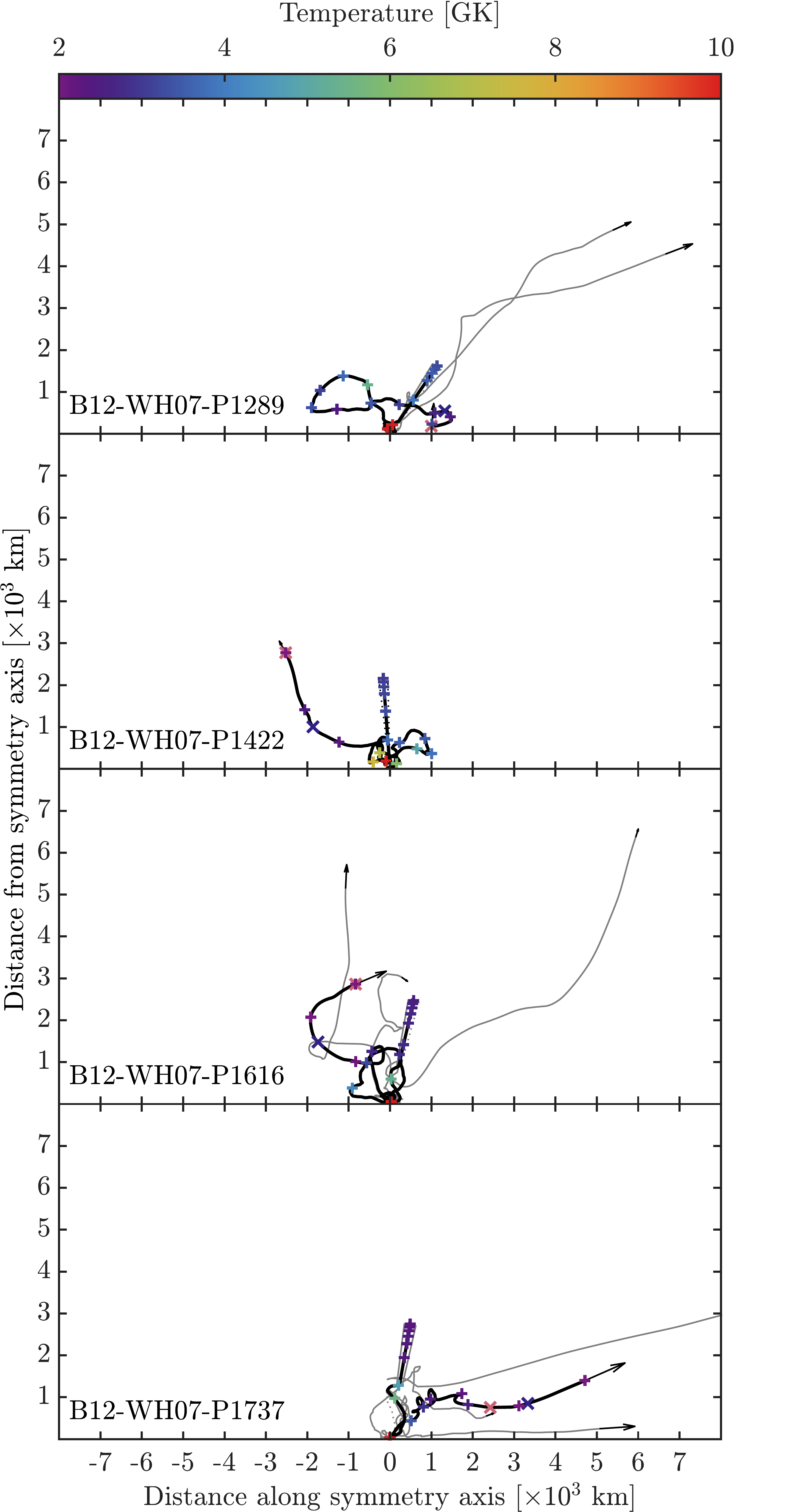}
	\caption{\label{fig:b12trajectory}
		Spatial trajectories (thick, black lines) for B12-WH07-P1289, B12-WH07-P1422, B12-WH07-P1616, and B12-WH07-P1737, with the initial neighboring tracer particles shown in gray and dotted if the particle does not represent part of the ejecta.
		Colored plus signs indicate particle temperatures in 100~ms increments from \tfinal.
		Red and blue crosses mark the points corresponding to \etaumax\ and \etaumin, respectively.
		The final velocity of each tracer particle, $\vec{v}(\tfinal)$, is represented by the arrows and scaled accordingly.
	}
\end{figure}

In order to determine the collective effect of single-particle extrapolation uncertainties, we post-process thermodynamic profiles generated by extrapolating from points in time in the last 150~ms of the simulation corresponding to the minimum and maximum estimates of the expansion timescales, \etaumin\ and \etaumax, respectively.
These changing timescales reflect the effects of ongoing hydrodynamic activity.
Of course, any extrapolation will fail to capture future hydrodynamic activity that deviates from true isentropic expansion, but changes over the last 150~ms are a reasonable proxy to estimate this uncertainty.
Defining a residual norm for the final composition,
\begin{equation}
	||r_{\detau}|| \equiv \frac{\sum_{i} |\detaui|}{\sum_{i} |\log_{10}(\sqrt{X_{i}(\etaumax)X_{i}(\etaumin)})|},
\label{eq:rdetau}
\end{equation}
where $\detaui = \log_{10}(X_{i}(\etaumax)/X_{i}(\etaumin))$ and $X_{i}$ is the mass fraction of species $i$, we are able to easily identify particles whose nuclear products are particularly susceptible to such activity.
Tracer particles with the largest extrapolation uncertainties, as measured by $||r_{\detau}||$, typically fall very close to the \PNS\ surface and reach peak temperatures in excess of 20~GK before being ejected at high speed (see Figure~\ref{fig:rdtau}).
The exposure of such particles to large neutrino fluxes near the neutrino-emitting surface, the \emph{neutrino-sphere}, is consistent with our premise that thermodynamic extrapolations initiated from $\Tfinal \lesssim 2$~GK are necessitated by the ongoing proton and neutron captures that typify neutrino-induced nucleosynthesis.
In Figure~\ref{fig:b12extrap}, we show differing estimates of the expansion timescale for four particles from B12-WH07 with the largest values of $||r_{\detau}||$: B12-WH07-P1289, B12-WH07-P1422, B12-WH07-P1616, and B12-WH07-P1737.
The differences are especially large for particle B12-WH07-P1289, which experiences a brief period of heating just prior to \tfinal, pushing the temperature above 3~GK and reengaging nuclear reactions that had otherwise ceased.
Hydrodynamically, this corresponds to the particle briefly halting its expansion as it retreats back towards the \PNS\ before being swept up in the ejecta once more (see Figure~\ref{fig:b12trajectory}).

However, this specific type of profile is very uncommon---particles with $||r_{\detau}|| > 0.1$ make up less than 1\% of the ejecta mass in B12-WH07.
In fact, B12-WH07-P1289 would have been a member of \Punbneg\ just 50~ms earlier.
For B12-WH07-P1422, B12-WH07-P1616, and B12-WH07-P1737, the differences in the thermodynamic extrapolations are less drastic but not insignificant.
In each of these cases, the differing estimates of the expansion timescale can be attributed to hydrodynamical flows which nudge the particles in question, but not to the extent that expansion ceases outright (see Figure~\ref{fig:b12trajectory}).
In so doing, a brief period of heating breaks the assumption of isentropic expansion.
For example, B12-WH07-P1422 exhibits significant differences in the final composition as a result of extrapolations that are quite similar but lagged by $\approx$150~ms.
Estimates of the expansion timescale in B12-WH07-P1737 differ by a factor of four despite the extrapolations initiating only $\approx$50~ms apart.
B12-WH07-P1616 represents a peculiar case; the alteration to its spatial trajectory, evident in Figure~\ref{fig:b12trajectory}, sends it on a collision course with the persisting equatorial downflow.
Though the particle is considered part of the ejecta at \tfinal, it is now part of a convective eddy and will inevitably soon transition to \Punbneg.
Not only does this highlight how nucleosynthesis predictions based on analytic extrapolations can fail to directly capture future hydrodynamic activity, but also elucidates the more subtle effect of different estimates of the expansion timescale.

Figure~\ref{fig:b12trajectory} illustrates the extent to which the spatial trajectories of these tracer particles can differ from those of their immediate neighbors, highlighting the highly asymmetric behavior which complicates multidimensional nucleosynthesis calculations.
As previously mentioned, a shared yet rare characteristic of the selected particles (B12-WH07-P1289, B12-WH07-P1422, B12-WH07-P1616, and B12-WH07-P1737) is their exposure to large neutrino fluxes near the neutrino-sphere.
As evidenced by their immediate neighbors, most tracer particles which descend deep into the gravitational well of the \PNS\ become bound and do not represent ejecta matter.

\begin{figure*}
	\centering
	\includegraphics[width=\textwidth,clip]{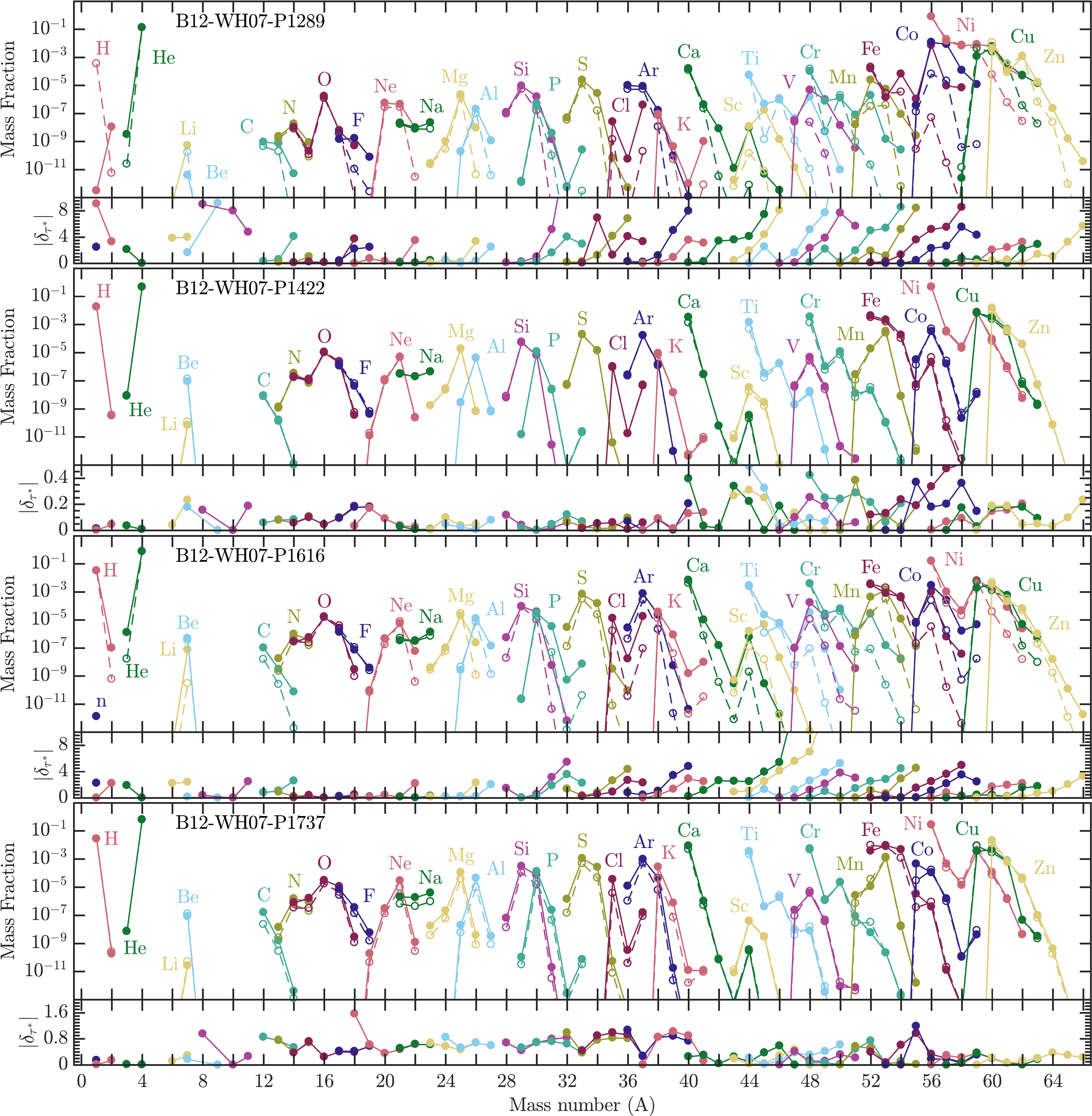}
	\caption{\label{fig:b12extrapmf}
		Top panels: Predicted ejecta mass fractions from post-processing each minimum/maximum pair of extrapolations for each particle shown in Figure~\ref{fig:b12extrap}.
		Bottom panels: Relative deviations of the composition between each pair of extrapolations plotted for each species $i$ as $\detaui \equiv \log_{10}(X_{i}(\etaumax)/X_{i}(\etaumin))$.
	}
\end{figure*}

At 100~s after bounce, the composition resulting from these varying extrapolations can be markedly different.
As shown in Figure~\ref{fig:b12extrapmf}, deviations from predicted trajectories after \tfinal\ can lead to non-trivial uncertainties in the composition for individual particles.

\begin{figure*}
	\centering
	\includegraphics[width=\textwidth,clip]{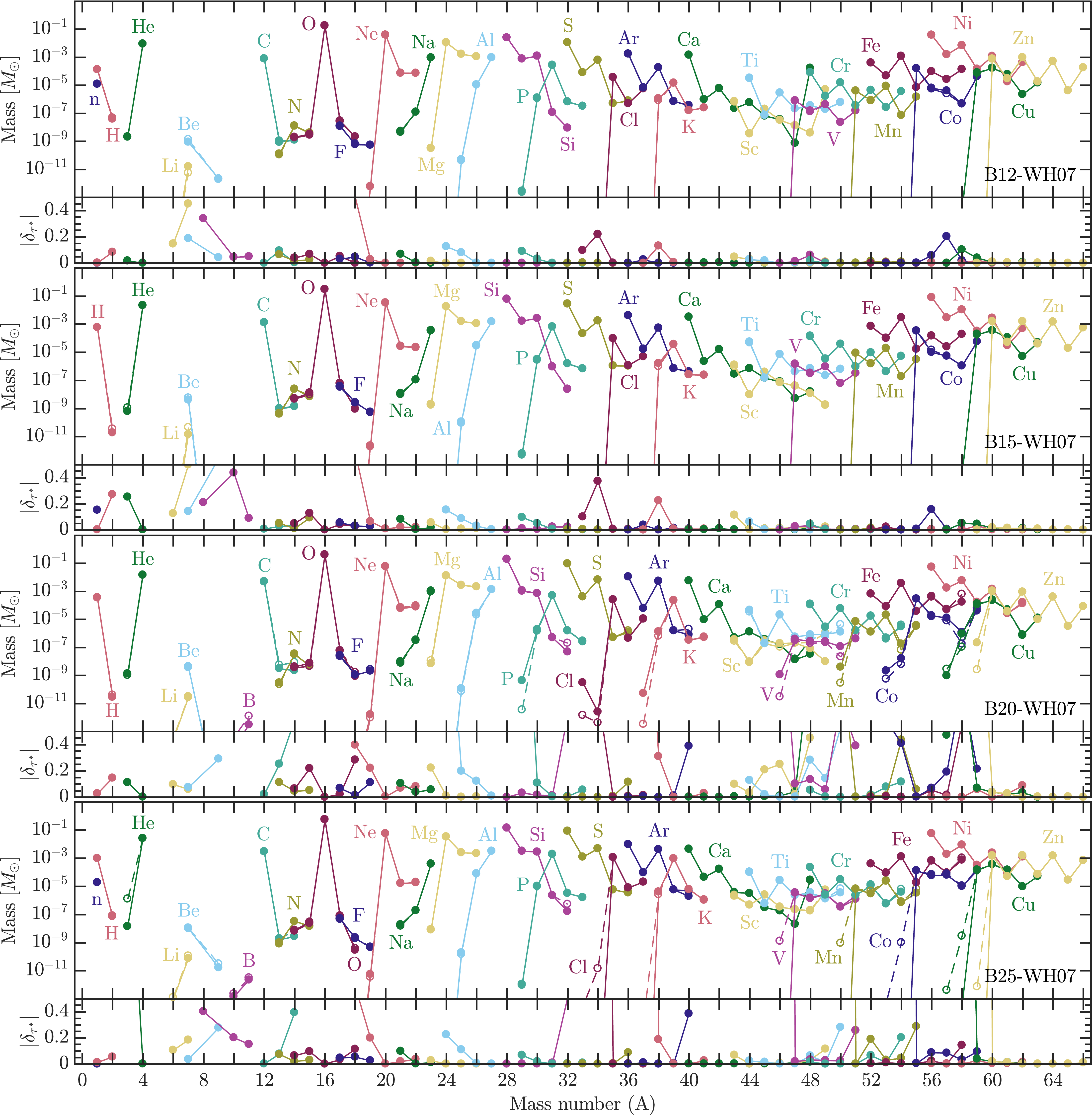}
	\caption{\label{fig:bseriesejectamassextrap}
		Top panels: \Mposi\ at 100~s after bounce and calculated for expansion timescales \etaumax\ (solid line) and \etaumin\ (dashed line) for particles in \Punbpos\ for each \SeriesB\ simulation.
		Bottom panels: Relative deviation of the composition between each pair of extrapolations plotted for each isotope $i$ as $\detaui \equiv \log_{10}(\Mposi(\etaumax)/\Mposi(\etaumin))$.
	}
\end{figure*}

The tracers discussed above were selected by their extreme uncertainties.
For most tracers, and therefore the total ensemble, the impact of the uncertainties is much less pronounced.
The extent to which this impacts the total ejecta mass for each nuclear species $i$, \Mposi, represented by \Punbpos\ (the particles for which we perform extrapolations) is shown for each \SeriesB\ model in Figure~\ref{fig:bseriesejectamassextrap}.
The relatively low temperatures of particles in $\Punbpos(\tfinal)$ (see Figure~\ref{fig:histtemp}) helps to limit the overall impact of extrapolation uncertainties to only a few nuclear products.
In general, $\detaui \equiv \log_{10}(\Mposi(\etaumax)/\Mposi(\etaumin))$ is larger for low-yield isotopes (e.g. \isotope{He}{3}, \isotope{C}{13}, \isotope{O}{18}, \ldots), wherein the extrapolation uncertainties for a small subset of tracer particles can have a greater effect.
We also identify trends of larger uncertainties for some isotopes that can't be entirely attributed to a small number of tracer particles (e.g. \isotope{K}{38}, \isotope{Sc}{43}, \isotope{Co}{56}, and \isotope{Co}{57}).
Uncertainties in \isotope{Ti}{44} production, important to supernova remnant observations, are small, but non-negligible ($0.79 \lesssim \Mposi(\etaumax)/\Mposi(\etaumin) \lesssim 0.95$; see Table~\ref{tab:uncertainties} for values).

Other isotopes are affected by variations in \etau, but their identities vary between models.
B20-WH07, for example, is uniquely characterized by a noticeable increase in \detaui\ for $12 \le A \le 30$.
B25-WH07 is also unique among the \SeriesB\ simulations in that the extrapolation error can be larger than 20\% for several of the most neutron-rich isotopes of elements for which $Z \ge 14$, including \isotope{Si}{32}, \isotope{S}{36}, \isotope{Ar}{40}, \isotope{Sc}{49}, \isotope{Ti}{50}, \isotope{V}{51}, \isotope{Cr}{53}, \isotope{Mn}{54}, \isotope{Fe}{58}, and \isotope{Co}{59}.

As a single measure of extrapolation uncertainties for a given model, we define a global residual using Equation~\ref{eq:rdetau} but replacing the mass fraction, $X_{i}$, with \Mposi.
Not surprisingly, $||r_{\detau}||$ increases with progenitor mass but is small relative to the other uncertainties (see Table~\ref{tab:uncertainties}).

\subsection{Spatial resolution}
\label{sec:dxresolution}

Any simulation, by its nature, replaces continuous thermodynamic variables with a discrete number of elements, either in space (Eulerian) or in mass (Lagrangian).
This discretization adds an uncertainty that is magnified if the number of elements is too small to capture the important features in sufficient detail.
For post-processing nucleosynthesis, the resolution in question is the number and distribution of the Lagrangian tracer particles used to gather the thermodynamic histories.

The initial distribution of tracer particles in our simulations guarantees that we are sampling the ejecta uniformly in mass.
Consequently, there are density-dependent variations in the spatial resolution, $\Delta\mathbf{r}$, of the ejecta.
If sufficiently large, these variations may limit the ability of Lagrangian tracer particles to sample the ejecta at resolutions which reproduce the nuclear burning conditions initially encountered in the simulation.
In particular, tracer particles tend to under-resolve regions of relatively low density.
While we could adjust the initial tracer mass to better sample low-density regions in the progenitor, our ability to do this for dynamically developing low-density regions is limited.
Past studies of the convergence properties of tracer particles in supernova nucleosynthesis \citep[see, e.g.,][]{SeRoFi10} have been limited to Type~Ia supernovae, which exhibit a narrower entropy range.

\begin{figure}
	\centering
	\includegraphics[width=\columnwidth,clip]{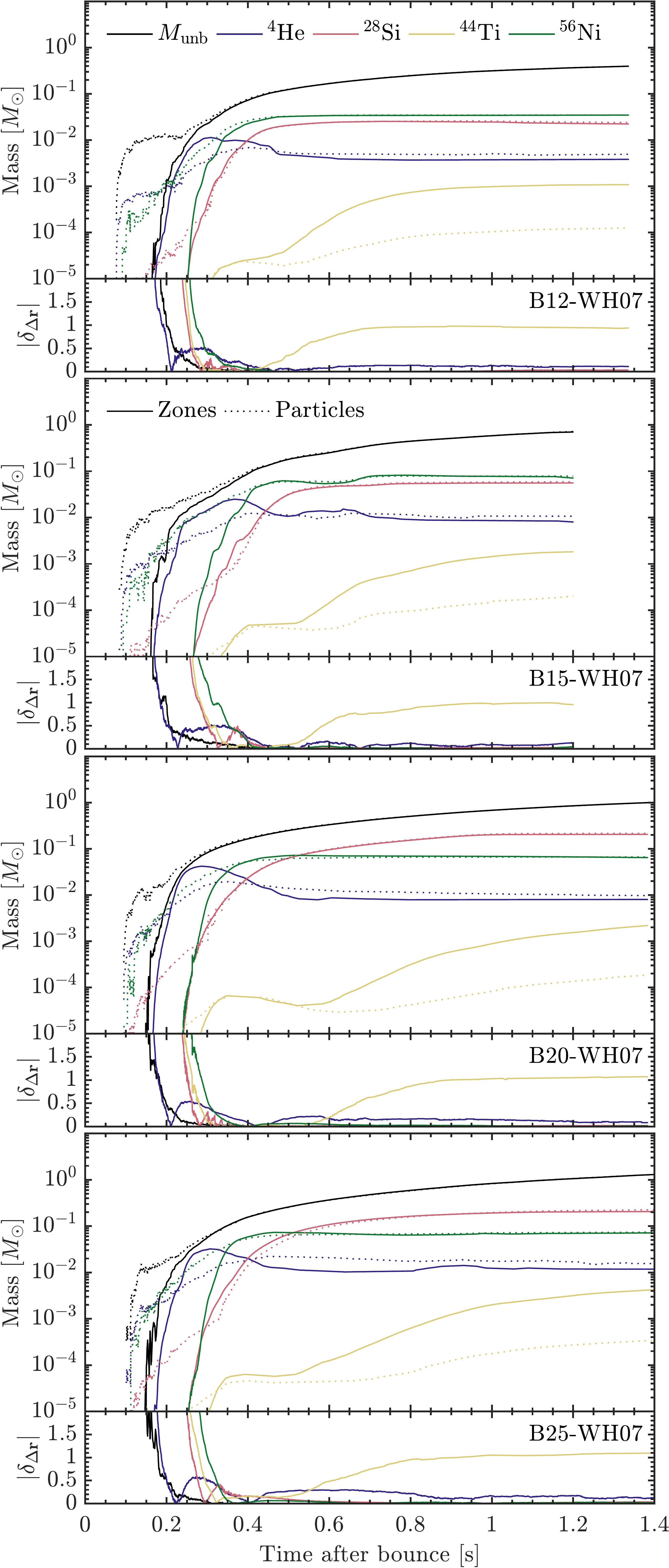}
	\caption{\label{fig:bseriesunb}
		Top panels: Total unbound mass of several \alp-network species as determined from the original simulation ($\Mchimi(t)$; solid lines) and post-processed using identical thermodynamic histories from Lagrangian tracer particles ($\MPPi(t)$; dotted lines) for each \SeriesB\ simulation.
		Bottom panels: Relative deviation from the original \emph{in situ} nucleosynthesis plotted as $\dDelri \equiv \log_{10}(\MPPi/\Mchimi)$.
	}
\end{figure}

We introduce here a novel way to explore the limitations of tracer post-processing by post-processing with a network identical to that used \emph{in situ} within \chimera.
Care is taken to also generate initial abundances, both for particles that reach NSE and those that do not, that are identical to the methods used within \chimera.
Thus, we have two representations of the nuclear composition that differ only in their effective resolution and the inability of the Lagrangian tracer particles to mix their composition.
The total ejected mass of each nuclear species $i$ from \emph{in situ} nucleosynthesis, \Mchimi, is calculated by integrating over all zones where $\edtot > 0$.
The ejected mass from the equivalent post-processing calculation, \MPPi, is calculated with the $\edtot > 0$ criterion applied to particles instead of zones.
The relative difference between \emph{in situ} and post-processing calculations for individual isotopes, $\dDelri \equiv \log_{10}(\MPPi/\Mchimi)$,
$\Mchimi(t)$, and $\MPPi(t)$ are shown throughout each \SeriesB\ simulation in Figure~\ref{fig:bseriesunb}.

There is general agreement between $\Mchimi(t)$ and $\MPPi(t)$ for most species.
\isotope{Si}{28} and \isotope{Ni}{56}, which can be taken as representation of their neighbors, show some early variance.
However, by \tfinal, and even by \texpl, these differences are only a few percent.
There is a consistent and stark discrepancy between the \emph{in situ} and post-processed total mass of \isotope{Ti}{44} (yellow lines) and, to a lesser extent, \isotope{He}{4}, found in all four models.
In B12-WH07, for example, the value from the \emph{in situ} calculation, $\MchimX{\isotope{Ti}{44}}(\tfinal) \approx 1.08\e{-3}~\msun$, is greater than that from post-processing, $\MPPX{\isotope{Ti}{44}}(\tfinal) \approx 1.24\e{-4}~\msun$, by nearly an order of magnitude.
To understand the origin of these inconsistencies, consider that the nuclei most affected are products of \alp-rich freeze-out occurring in low-density, expanding ejecta and are, therefore, most susceptible to inadequate tracer particle spatial resolution.
The deficit of \alp-rich freeze-out products in the post-processing results cannot be attributed to a lack of mixing therein, as mixing would dilute the \alp-richness of this ejecta and, therefore, further reduce the production of these nuclei.

\begin{figure}
	\centering
	\includegraphics[width=\columnwidth,clip]{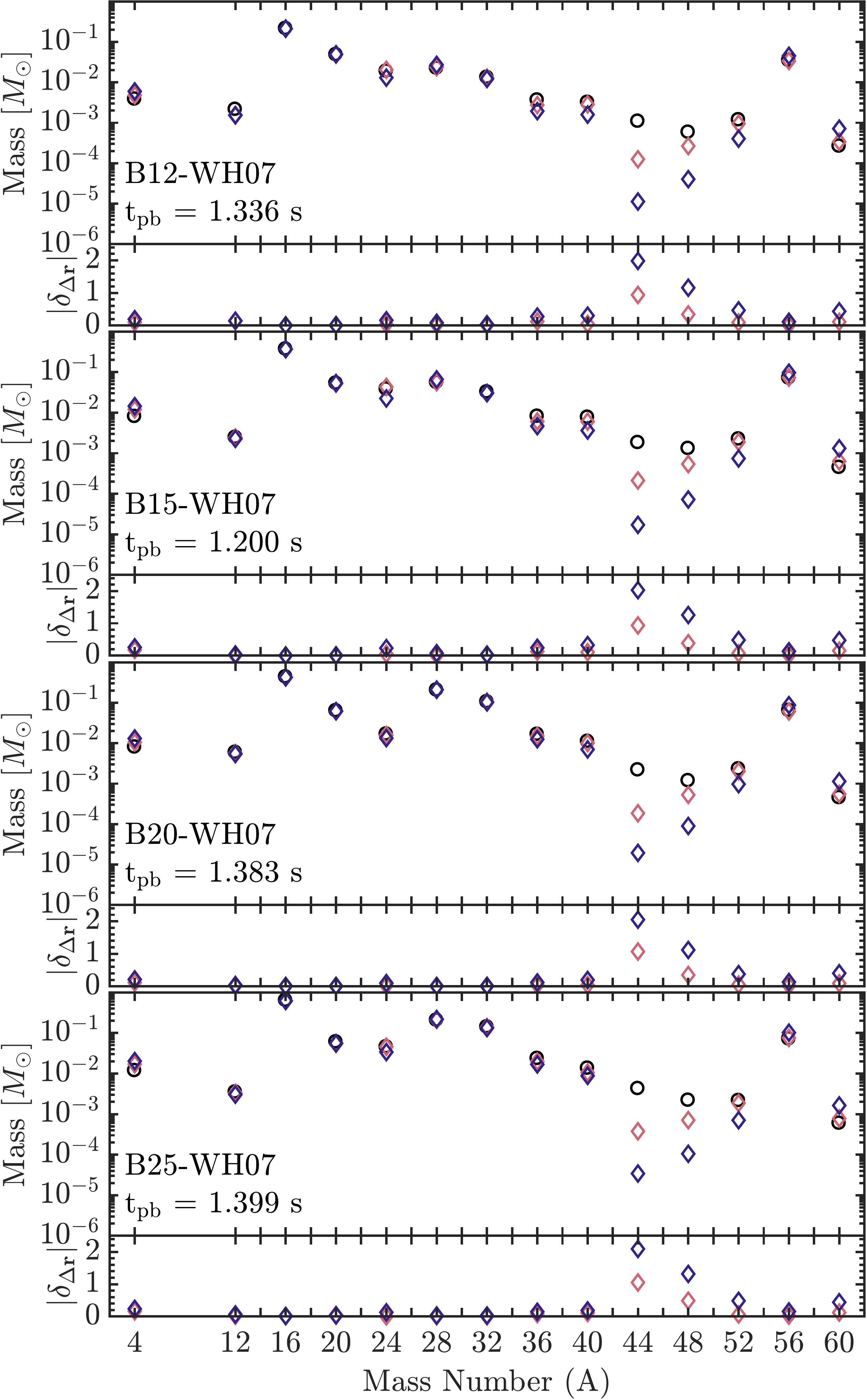}
	\caption{\label{fig:bseriesunbfinal}
		Top panels: \Mchimi\ (black circles) and \MPPi\ (diamonds) at the end of each simulation and transitioned out of NSE using $\TNSE(\rho)$ from Equation~\ref{eq:tnse}.
		The post-processed nucleosynthesis results are shown for two different networks: the \alp-network used in the simulation (red) and a 150-species network (blue).
		Bottom panels: Deviations from the original \emph{in situ} nucleosynthesis plotted for each \alp-network isotope $i$ as $\dDelri \equiv \log_{10}(\MPPi/\Mchimi)$.
	}
\end{figure}

To provide context regarding the magnitude of this disagreement, we compare $\Mchimi(\tfinal)$ and $\MPPi(\tfinal)$ to post-processing calculations performed with a more realistic 150-species network in Figure~\ref{fig:bseriesunbfinal}.
In each model, the unbound \isotope{Ni}{56} mass is relatively unaffected, but there are significant differences in the total unbound mass for \isotope{Ti}{44}, \isotope{Cr}{48}, \isotope{Fe}{52}, and \isotope{Zn}{60}.
These differences can be largely attributed to the availability of additional reaction pathways during explosive burning, particularly those involving $(n,\gamma)$ and $(p,\gamma)$ reactions \citep{HiTh96,TiHoWo00,MaTiHu10}.
The effect of spatial resolution on the production of \isotope{Ti}{44} is of roughly the same order, but in the opposing direction, as that seen when switching from the \alp-network to \snnet.
Without an \emph{in situ} large network simulation, we are unable to fully quantify how spatial resolution of the tracer particles may impact nucleosynthesis with realistic nuclear networks.
However, this work suggests that much higher tracer particle spatial resolution, and ultimately, large, \emph{in situ} nuclear networks are needed to correctly calculate \alp-rich freeze-out.

\begin{figure}
	\centering
	\includegraphics[width=\columnwidth,clip]{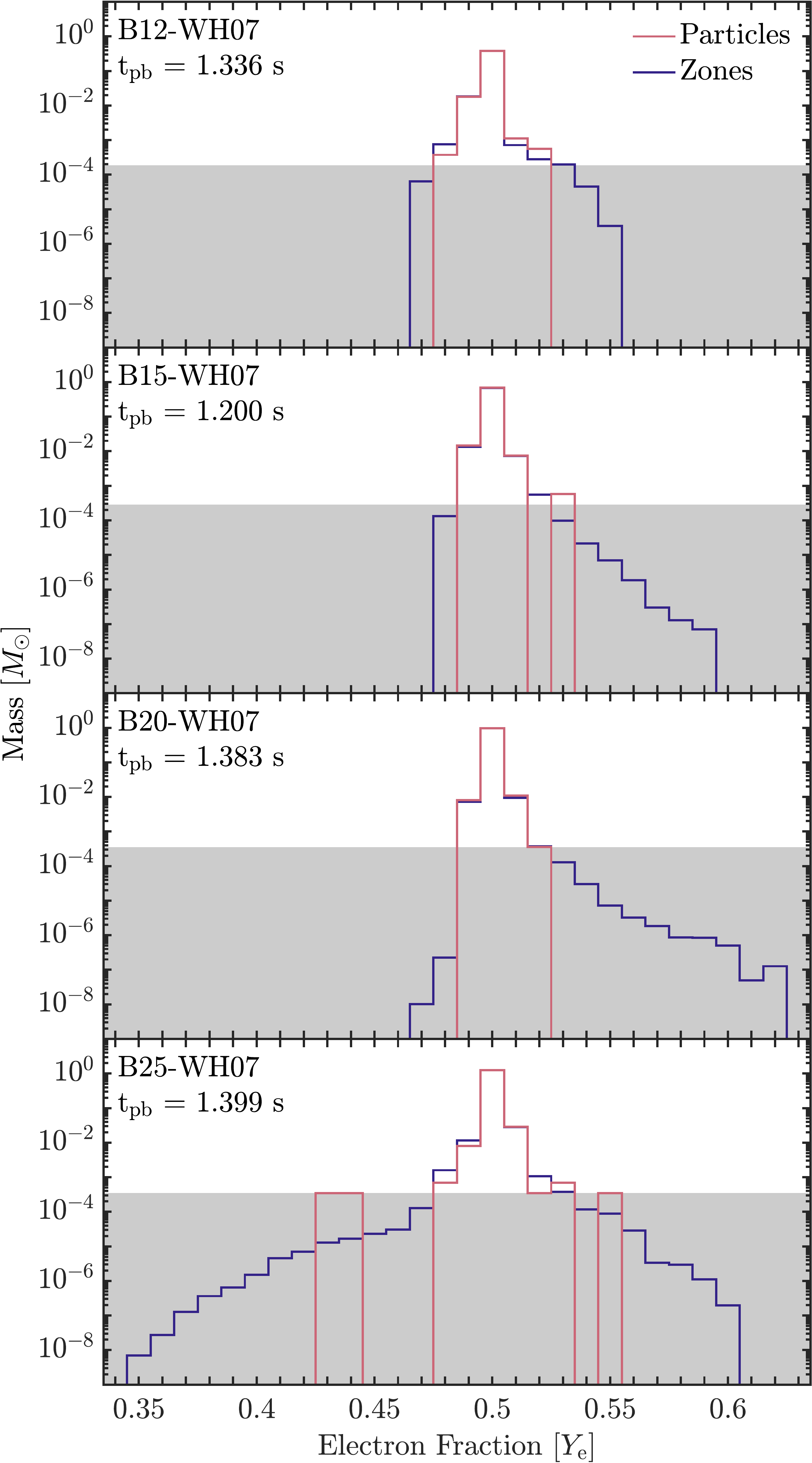}
	\caption{\label{fig:histye}
		Mass histograms of $\Munb(\tfinal)$, with bin sizes of $\Delta \Ye(\tfinal) = 0.01$ for \SeriesB\ models and calculated using both unbound particle data (red) and grid data (blue).
		The mass represented by one Lagrangian tracer particle is given by the shaded region.
	}
\end{figure}

The susceptibility of post-processing to inadequate spatial resolution can also be broadly characterized by the mass distribution of $\Ye(\tfinal)$ for unbound matter (see Figure~\ref{fig:histye}).
In these mass histograms, calculated from both individual zone data (blue) and tracer particle data (red), it becomes clear that the masses represented by a single tracer particle (shaded region) in the \SeriesB\ models fail to adequately resolve ejecta outside of $0.49 < \Ye(\tfinal) < 0.51$.
While this constitutes by far the majority of the matter, it neglects some of the most interesting nucleosynthesis.
For B12-WH07 and, to a large extent, B15-WH07, increasing the number of tracer particles by an order of magnitude would serve to capture much more of the \Ye\ mass distribution and is a relatively easy solution to implement.
Applying a similar approach to B20-WH07 and B25-WH07 would require an untenable number of particles to effectively capture the long tails of the distribution.

As in Equation~\ref{eq:rdetau}, we define
\begin{equation}
	||r_{\dDelr}|| \equiv \frac{\sum_{i} |\dDelri|}{\sum_{i} |\log_{10}(\sqrt{\Mchimi(\tfinal)\MPPi(\tfinal)})|}.
\label{eq:rdDelr}
\end{equation}
From this measure, it becomes clear that uncertainties stemming from spatial resolution, even though they apply to only a small subset of species, are the largest, single source of nucleosynthesis error in the \SeriesB\ models (see Table~\ref{tab:uncertainties}).

\subsection{NSE transition and network size}
\label{sec:nsetransition}

Compositional evolution via the nuclear reaction network, both within the simulation and as a post-processing calculation, is an initial-value problem, with the initial values provided by one of two different methods (Section~\ref{sec:networks}).
For particles which have never reached NSE, the initial composition of the progenitor is mapped onto the nuclear network used for the nuclear evolution, providing the initial abundances.
For particles which have reached NSE, the initial abundances from a NSE calculation can be mapped onto the network.
\chimera's treatment of the transition of matter into NSE (Section~\ref{sec:nucleosynthesis}) is comparable to (or in some cases, better than) that used in other CCSN codes of similar capability---e.g., \castro\ \citep{AlBeBe10,ZhHoAl11,ZhHoAl13}, \prometheusvertex\ \citep{BuRaJa06,MaJa09}, \coconutvertex\ \citep{MuJaMa12}, \zelmani\ \citep{OtOcPe11,OtAbMo13}, and \zidsa\ \citep{SuKoTa10,SuTaKo13}.
The transition condition is motivated by the temperatures and densities at which complete silicon burning would occur within the current global timestep.
For temperatures above this threshold, the use of the nuclear network is superfluous, as the network will achieve NSE every timestep.
For simplicity, the transition out of NSE occurs when the temperature drops below this NSE condition (Equation~\ref{eq:tnse} for \chimera).
However, for the rapidly changing conditions in expanding CCSN matter, the assumption of NSE has been shown to break down when the temperature falls below 6~GK \citep{MeKrCl98}.

\begin{figure}
	\centering
	\includegraphics[width=\columnwidth,clip]{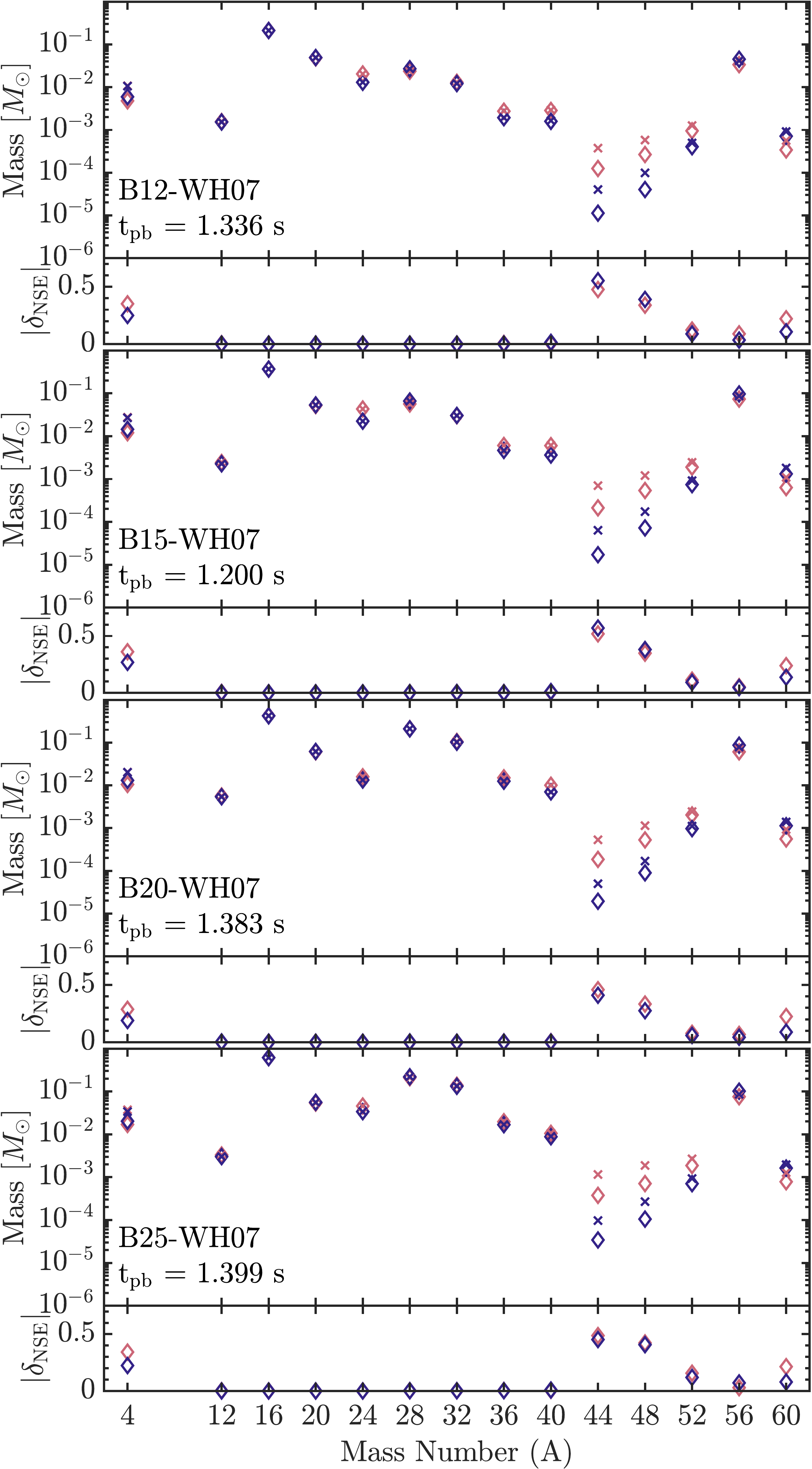}
	\caption{\label{fig:bseriesunb_tnsefinal}
		Top panels: $\MPPi(\tfinal)$ transitioned out of NSE using $\TNSE(\rho)$ (diamonds) and $\TNSE = 8$~GK (crosses).
		The post-processed nucleosynthesis results are shown in the top panels for two different networks: the \alp-network used in the simulation (red) and a 150-species network (blue).
		Bottom panels: Relative deviation of composition between different values of \TNSE\ plotted for each species $i$ as $\dNSEi = \log_{10}(\MPPi(\TNSE(\rho))/\MPPi(\TNSE = 8~\mathrm{GK}))$.
	}
\end{figure}

This leaves a dilemma for this or any similar post-processing study.
Is it better to be consistent with the NSE-to-network transition used within the supernova simulation, or should an earlier (higher temperature) transition out of NSE be adopted for the network?
As a test of the NSE transition criteria in the \SeriesB\ simulations, we post-process the nucleosynthesis using the same \alp-network used with the simulations and vary the conditions at which the transition to nuclear burning from the NSE composition occurs.
In Figure~\ref{fig:bseriesunb_tnsefinal}, we compare the unbound masses of individual isotopes, transitioned out of NSE using either $\TNSE(\rho)$, as defined in Equation~\ref{eq:tnse}, or $\TNSE = 8$~GK and show the relative deviation for each species $i$ as $\dNSEi \equiv \log_{10}(\MPPi(\TNSE(\rho))/\MPPi(\TNSE = 8~\mathrm{GK}))$.
The \emph{in situ} calculation transitions out of NSE at a temperature $\approx$2--3~GK lower than the tested value of 8~GK.
This reduces the \alp-richness of the eventual freeze-out by maintaining NSE longer than it would physically.
As a result, we see a significant shift in the masses of \isotope{He}{4}, \isotope{Ti}{44}, \isotope{Cr}{48}, and \isotope{Zn}{60} in each model (e.g., $\MPPX{\isotope{Ti}{44}}(\TNSE=8~\mathrm{GK})/\MPPX{\isotope{Ti}{44}}(\TNSE(\rho)) \approx 3.0$), highlighting the failure of $\TNSE(\rho)$ to fully capture \alp-rich freeze-out in the ejecta.
This argues that for a stricter set of criteria for the breakdown of NSE should be adopted for use in post-processing than the $\approx$5--6~GK commonly used within models of the CCSN mechanism so that one may properly achieve \alp-rich freeze-out.
Furthermore, the transition used within the models themselves should be questioned if direct nucleosynthesis results are to be used to constrain models.
We define
\begin{equation}
	||r_{\dNSE}|| \equiv \frac{\sum_{i} |\dNSEi|}{\sum_{i} |\log_{10}(\sqrt{\MPPi(\TNSE(\rho)) \MPPi(\TNSE = 8~\mathrm{GK})})|}
\label{eq:rdNSE}
\end{equation}
for direct comparison with the uncertainties defined by Equation~\ref{eq:rdetau} and Equation~\ref{eq:rdDelr} (see Table~\ref{tab:uncertainties}).

The same post-processing calculations using \chimera's \alp-network also provide insight into the limitations of this network on the nucleosynthesis in the simulation.
Limitations in tracer particle spatial resolution significantly constrain our ability to capture \alp-rich freeze-out in post-processing calculations (Section~\ref{sec:dxresolution}), but this uncertainty is mildly abated by improved treatment of the NSE transition, which enhances production of \alp-rich freeze-out species (see Figure~\ref{fig:bseriesunb_tnsefinal}).
This effect is visible for both \snnet\ and the \alp-network, and $\dNSEi(\alp)$ is nearly identical to $\dNSEi(\snnet)$.
Differences in ejected mass of individual isotopes incurred from nuclear network size can vary (compare blue diamonds to red diamonds in Figures~\ref{fig:bseriesunb_tnsefinal}); the extent of which can be as high as an order of magnitude for \isotope{Ti}{44}.
However, for species like \isotope{Ni}{56}, the effect is smaller ($\approx$20\%).
With this uncertainty in mind, we have reasonable confidence quoting the production of Ni and species from O--Ca from the \emph{in situ} \alp-network, but the values for the products of \alp-rich freeze-out require the use of a larger \emph{in situ} network.

\subsection{Time resolution}
\label{sec:dtresolution}

The convective and turbulent nature of CCSNe make it highly probable for tracer particle thermodynamic states to change very rapidly.
To capture this detail, in the B-series models, \chimera\ recorded the temporal history of each particle independently whenever the temperature changed by 0.1\% from the last record and, likewise, limited changes in density and integrated neutrino number flux, \phinu, to 1\%.
These sampling criteria ensure that all significant features in the temporal histories of the tracer particles are captured.
However, this high temporal cadence limits the number of tracers that can be evolved if the total cost of the simulation is not to be limited by tracer I/O.
Independent sampling of individual tracers for the hundreds-of-thousands of tracer particles necessary to adequately sample a 3D simulation poses an even heavier and perhaps untenable load on I/O.
One alternative to this approach is to record the thermodynamic states of all tracer particles at fixed time intervals along with the typical checkpoint data of the simulation.
While this addresses the logistical concerns of managing a large amount of data, it runs the risk of insufficiently sampling the histories of individual particles.
The biggest danger of less frequent sampling is the underestimate of local maxima in temperature and density in the thermodynamic history.
The exponential temperature dependence of thermonuclear reactions can potentially greatly magnify the impact of under-sampling in the local peak conditions.
The balanced reactions which maintain NSE, particularly those linking \isotope{He}{4} to \isotope{C}{12}, are especially susceptible to peak temperature and density conditions \citep{MeKrCl98}.
Thermodynamic histories which fail to capture this peak behavior due to inadequate time resolution can thereby misestimate the conditions of \alp-rich freeze-out.

\begin{figure}
	\centering
	\includegraphics[width=\columnwidth]{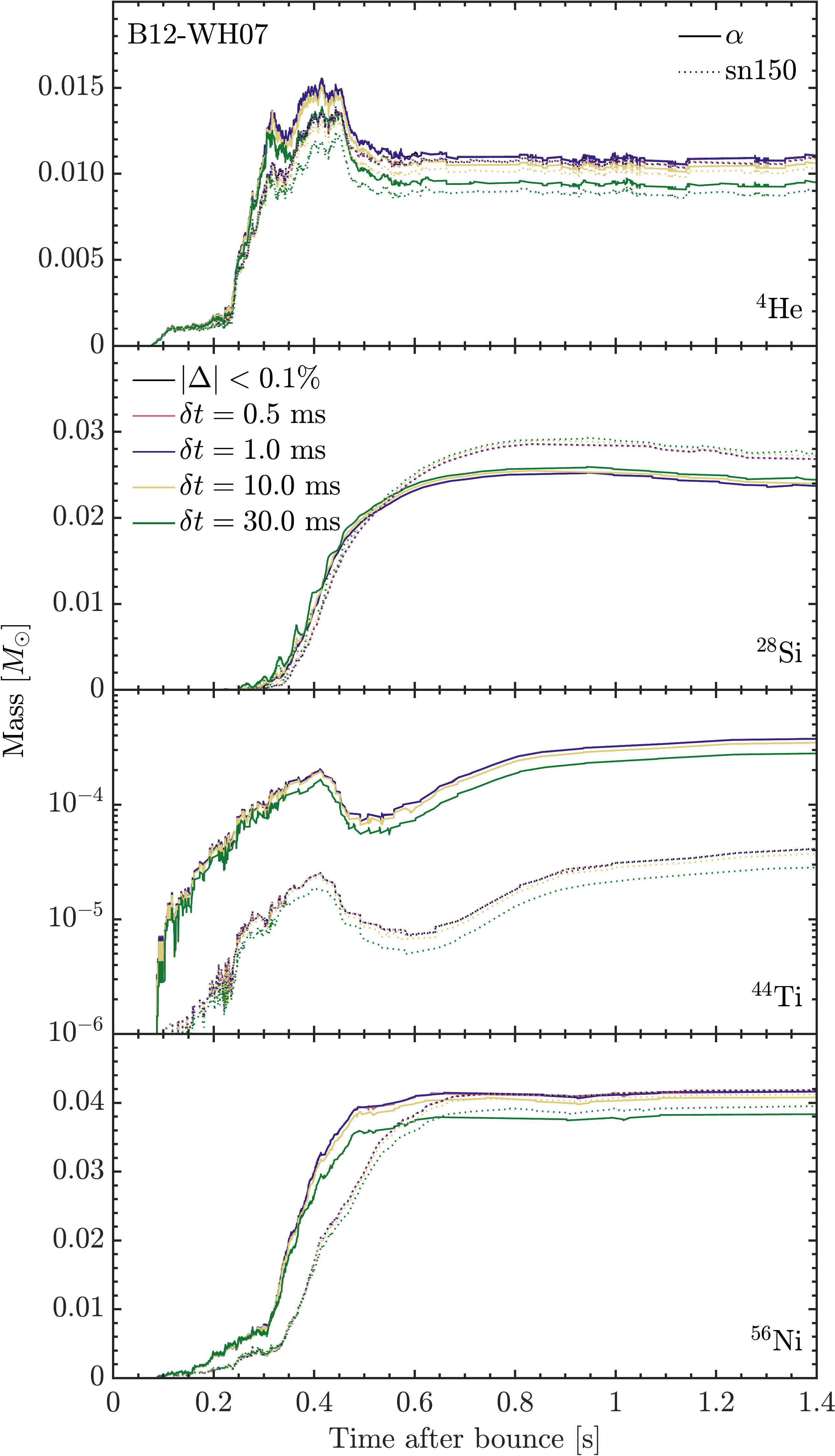}
	\caption{\label{fig:b12unbdt}
		B12-WH07 total unbound mass, \Munb, for \isotope{He}{4}, \isotope{Si}{28}, \isotope{Ti}{44}, and \isotope{Ni}{56} from post-processing calculations using both the \alp-network (solid lines) and \snnet\ (dotted lines) and performed with thermodynamic profiles generated using different fixed sampling intervals, $\delta t$, and a dynamic time interval chosen such that $|\Delta| < 0.1\%$, where $\Delta = \max(|\Delta T|,|\Delta \rho|/10,|\Delta \phinu|/10)$.
	}
\end{figure}

We quantify this effect and other sampling-related concerns in tracer particle histories by post-processing thermodynamic profiles down-sampled to fixed time intervals.
We compare the resulting composition profile to that from the dynamic time interval criteria described above for \isotope{He}{4}, \isotope{Si}{28}, \isotope{Ti}{44}, and \isotope{Ni}{56} for both the \alp-network and \snnet\ in Figure~\ref{fig:b12unbdt} for B12-WH07.

We find that a fixed rate of sampling the thermodynamic state reproduces post-processed abundances from the dynamic time interval criteria to within 0.5\% so long as the fixed interval, $\delta t$, does not exceed 1~ms.
For $\delta t > 1$~ms, post-processing calculations fail to capture the important fluctuations in density and temperature, particularly those that define the conditions of \alp-rich freeze-out.
The subsequent errors incurred in the total unbound mass is independent of the reaction network size, and \isotope{Ti}{44} is most affected by poor time resolution ($\approx$10\% error for $\delta t = 10$~ms and $\approx$30\% error for $\delta t = 30$~ms).

\section{Gauging the completeness of the thermonuclear evolution}
\label{sec:doneness}

The goal of any calculation of CCSN nucleosynthesis is the final distribution of the ejecta in composition and velocity.
The evolution to this final state can be divided into four rough stages:
(1) the development of the explosion; (2) the development of the mass-cut that separates ejecta from \PNS; (3) the completion of the nucleosynthesis as all ejecta expands and cools sufficiently for nuclear reactions to cease; and (4) the development of the final velocity distribution for the outgoing ejecta as the shock passes through the stellar envelope.
In practice, these four stages are not so clearly demarcated as this list implies, and the overlap of these stages is exacerbated in multidimensional simulations.

For investigations primarily focused on the development of the explosion, there is a strong temptation to conserve computational resources by stopping the simulation once the successful revival of the supernova shock seems guaranteed or, at the latest, as the growth of the explosion energy levels off.
From the nucleosynthesis perspective, such an abbreviation of the simulations certainly prevents examination of the final stage in the prior list; however, it is unclear how the intermediate stages may be affected.
In this section, we attempt to quantify the impact of early termination of the simulation in terms of the uncertainties discussed in Section~\ref{sec:uncertainties}.

From the global analysis of the nuclear composition discussed in \citet[][see Figure 22]{BrLeHi16}, it is clear that the nuclear composition is changing dramatically even 600~ms after bounce in all four models we consider here.
This corresponds to the epoch when the explosion energy in these models begins to level off.
We define this time as \texpl\ and estimate $\texpl = 600$~ms for B12-WH07 and $\texpl = 800$~ms for the other \SeriesB\ simulations \citep[see Figure~12a in][]{BrLeHi16}.
The decline in the growth of the explosion energy is concurrent with a similar decrease in the growth of the unbound mass (\Munb; see Figure~\ref{fig:bseriesmass}); however, using $\eps(t)$ in these figures as a proxy for the evolution of the multidimensional mass-cut, we see that the models at \texpl\ are much less mature than at \tfinal.
The disparity between $\eps(\texpl)$ and $\epshat(\texpl)$ clearly demonstrates the particles representative of ejecta are in flux, further supported by the growth in \Mbound\ seen after \texpl\ in each model as the shock moves outward before leveling off near \tfinal.

\begin{figure*}
	\centering
	\includegraphics[width=\textwidth,clip]{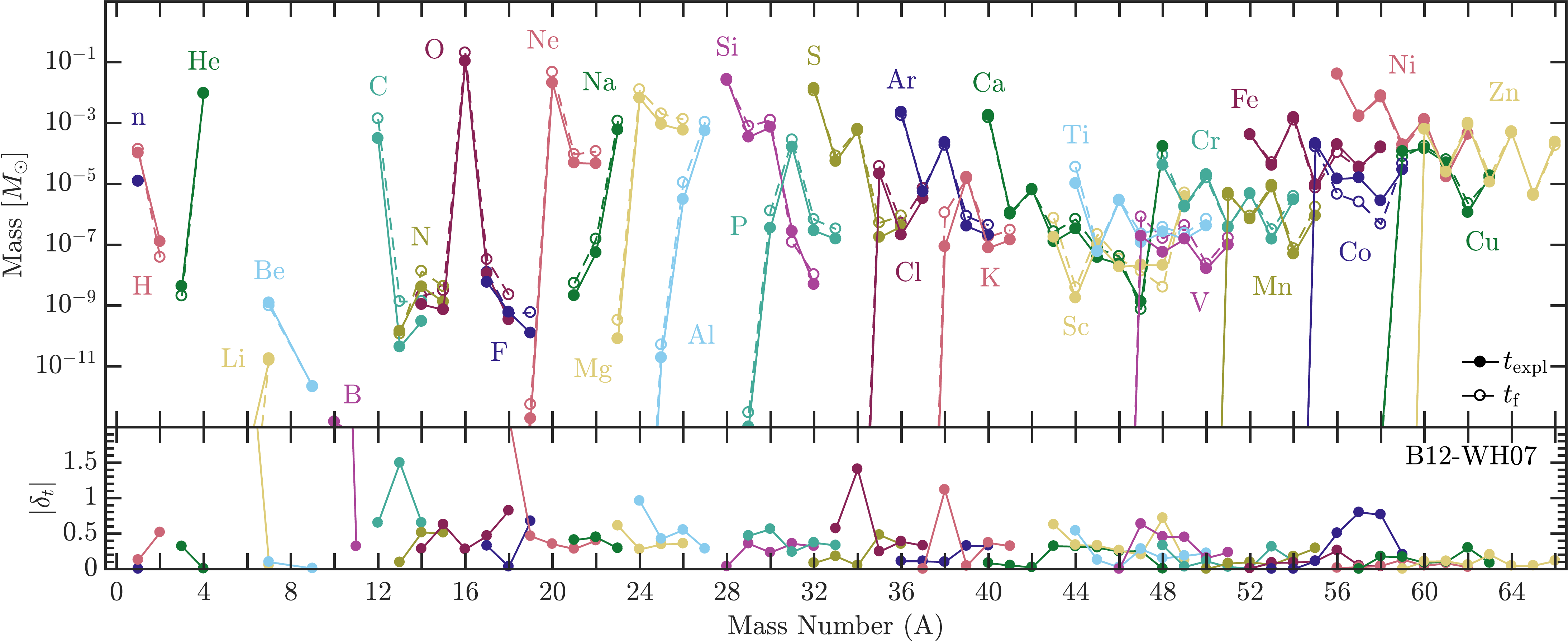}
	\caption{\label{fig:b12extrapexpl}
		Top panel: \Mposi\ of B12-WH07 at 100~s after bounce, calculated from \texpl\ (solid lines) and \tfinal\ (dashed lines).
		Bottom panel: Relative deviation of the composition between the different end times plotted for each isotope $i$ as $\dti \equiv \log_{10}(\Mposi(\texpl)/\Mposi(\tfinal))$.
	}
\end{figure*}

Such early truncation of the simulation often occurs with much of the ejecta at final temperatures above 3~GK (see Figure~\ref{fig:histtemp}; yellow) and fails to capture all stages of explosive nucleosynthesis \emph{in situ} when coupling to the hydrodynamics is important.
In this case, investigation of the nucleosynthesis will much more heavily rely on extrapolations of the thermodynamic history, described in detail in Section~\ref{sec:trajectories}.
Coming, as it does, at a point the supernova's evolution where hydrodynamics remains very active, this extrapolation is fraught with uncertainty.
As a measure of this, we compare predictions of the nucleosynthetic yields determined from \texpl\ and \tfinal\ at 100~s in Figure~\ref{fig:b12extrapexpl}.
Whereas uncertainties relating to thermodynamic extrapolation at \tfinal\ (see Section~\ref{sec:extrap}) are small relative to other uncertainties, the magnitude of variations between $\Mposi(\texpl)$ and $\Mposi(\tfinal)$ for $A \ge 12$, which includes the effects of both a different multidimensional mass-cut and contrasting expansion timescales, can exceed an order of magnitude, with a factor of three being quite common.
Interestingly, with the exception of \isotope{Ti}{44}, the \alp-nuclei from silicon to nickel are relatively unaffected (e.g. $\MposX{\isotope{Ni}{56}}(\texpl)/\MposX{\isotope{Ni}{56}}(\tfinal)=1.033$).
The production of these nuclei is largely tied to complete, explosive silicon burning ignited by the passage of the shock \citep[see Table III in][]{WoHeWe02}.
As seen in Figure~\ref{fig:histtemp}, by \texpl, the shock is no longer heating newly swept up matter sufficiently ($T\gtrsim5$~GK).

\section{Summary}
\label{sec:summary}

In this paper, we have discussed some of the uncertainties which complicate post-processing nucleosynthesis calculations from \emph{ab initio} multidimensional CCSN simulations evolved beyond the initial stages of explosive nuclear burning but not yet to conditions for nuclear reactions to cease.
We provide specific examples of how these uncertainties impact nucleosynthesis predictions for the four axisymmetric models of \citet{BrLeHi16}.
A detailed exposition of nucleosynthesis for these models using very large networks constitutes the subject of a forthcoming publication \citep[][in prep]{HaHiCh17}.

Our results can be summarized as follows:
\begin{enumerate}
	\item\label{itm:1} Even after 1.2--1.4~s of post-bounce evolution and with asymptotic explosion energies, the multidimensional mass-cut remains unresolved in each model, impacting the production of nuclear species in borderline ejecta near ongoing accretion flows.
	The consequences of this are particularly pronounced in B25-WH07, wherein the ultimate fate of $\approx$0.2~\msun\ of a total $\approx$1.3~\msun\ of gravitationally-unbound matter is indeterminate.
	The state of the multidimensional mass-cut is even more dire if the simulations are truncated at 600--800~ms after bounce, after the explosion energy has begun to level off but before explosive nucleosynthesis has completed.
	\item\label{itm:2} Despite temperatures below the threshold for explosive nuclear burning, a result of the extended running times, local hydrodynamic deviations from isentropic expansion continue to play a non-trivial role on secondary nuclear processes and neutrino-induced nucleosynthesis by altering the expansion timescale estimate necessary for extrapolation to freeze-out at $T \approx 0.5$~GK.
	\item\label{itm:3} The ability of Lagrangian tracer particles to effectively reproduce the \emph{in situ} nuclear burning conditions of the \SeriesB\ simulations is significantly reduced in regions of \alp-rich freeze-out.
	In each of the four models, \isotope{Ti}{44} is consistently under-produced.
	The magnitude of this effect is similar, but in the opposing direction, to that of replacing the \alp-network used in the post-processing calculation with a more realistic 150-species network.
	While this effect targets only those species that result from \alp-rich freeze-out, it has the largest global effect of the uncertainties tested.
	Ideally, large \emph{in situ} networks are the best method to address this issue.
	Failing that, much higher numbers of tracer particles are needed.
	\item\label{itm:4} Furthermore, we argue for a stricter set of criteria for transitioning out of NSE than the commonly used $\TNSE \approx 5\textrm{--}6$~GK within models of the CCSN mechanism.
	While sufficient for the transition of matter into NSE, this criteria fails to capture the process of \alp-rich freeze-out crucial to the production of \isotope{Ti}{44}.
	\item\label{itm:5} We find that recording the thermodynamic conditions of all tracer particles at a fixed time interval $\delta t \le 1$~ms to be a viable alternative to independently tracking the detailed history of each particle, limited by deviations in the nuclear burning inputs (i.e. $\rho$, $T$, and \phinu).
\end{enumerate}

In light of these findings, we have modified \chimera, when possible, in an attempt to reduce these uncertainties for future models.
Regarding Item~\ref{itm:3}, ongoing models improve the spatial resolution of tracer particle sampling by increasing the number of tracer particles by roughly one order of magnitude.
The larger burden this would place on the file system is sufficiently alleviated with fixed time interval I/O for tracer particles.
Our results in Section~\ref{sec:dtresolution} (Item~\ref{itm:5}) give us confidence in this approach.
Lastly, taking our own advice regarding Item~\ref{itm:4} (Section~\ref{sec:nsetransition}), we have implemented a framework in \chimera\ to use arbitrary criteria for the NSE transition.
When combined with quasi-statistical equilibrium methods \citep{HiPaFr07,PaHiTh08b}, \chimera\ will be capable of seamlessly evolving nuclear burning networks through the NSE transition.

In theory, both the indeterminate mass-cut and expansion timescale uncertainties could be reduced by extending the simulation to freeze-out.
However, given the inadequate spatial resolution of the tracer particles and an inherent limitation in the accuracy of the rate of nuclear energy released by the smaller network within the hydrodynamics, we cannot rely entirely on post-processing methods to obtain an accurate representation of the nucleosynthesis.
Since the nucleosynthesis depends on the thermodynamic conditions and, consequently, nuclear energy generation, a feedback exists that cannot be captured with post-processing, significantly affecting the abundances of species such as \isotope{Ti}{44}, \isotope{Fe}{57}, \isotope{Ni}{58}, and \isotope{Zn}{60} \citep{WoWe95}.
Improving upon the existing \emph{in situ} \alp-network with a more realistic 150-species nuclear network capable of properly tracking neutronization and energy release via particle captures is an important step towards resolving this problem and is the subject of ongoing work.

The research presented here was supported by the U.S. Department of Energy, Office of Science, Offices of Nuclear Physics and Advanced Scientific Computing Research, the NASA Astrophysics Theory and Fundamental Physics Program (NNH11AQ72I) and the National Science Foundation Nuclear Theory Program (PHY-1516197).
The simulations described herein were performed via NSF TeraGrid resources provided by the National Institute for Computational Sciences under grant number TG-MCA08X010; resources of the National Energy Research Scientific Computing Center, supported by the U.S. DOE Office of Science under Contract No.~DE-AC02-05CH11231; and an award of computer time from the Innovative and Novel Computational Impact on Theory and Experiment (INCITE) program at the Oak Ridge National Leadership Computing Facility, supported by the U.S. DOE Office of Science under Contract No.~DE-AC05-00OR22725.

\bibliographystyle{apj}

\end{document}